\documentclass[a4paper,11pt]{article}
\pdfoutput=1

\usepackage{jcappub} 
\usepackage{aas_macros}
\usepackage{placeins}
\usepackage{comment}
\usepackage{graphicx}
\usepackage{multirow}
\usepackage{amsmath,amssymb,amsfonts}
\usepackage{amsthm}
\usepackage{mathrsfs}
\usepackage[title]{appendix}
\usepackage{xcolor}
\usepackage{textcomp}
\usepackage{manyfoot}
\usepackage{booktabs}
\usepackage{algorithm}
\usepackage{algorithmicx}
\usepackage{algpseudocode}
\usepackage{listings}
\usepackage{subfigure}

\usepackage{tikz}
\usetikzlibrary{calc,positioning,fit,arrows.meta}
\usepackage{float}
\pdfoutput=1 
\usepackage[T1]{fontenc}

\theoremstyle{thmstyleone}

\theoremstyle{thmstyletwo}

\theoremstyle{thmstylethree}

\newcommand{\vx}{\mathbf{x}}

\newcommand{\vg}{\mathbf{g}}
\newcommand{\vu}{\mathbf{u}}
\newcommand{\vy}{\mathbf{y}}
\newcommand{\vq}{\mathbf{q}}
\newcommand{\vp}{\mathbf{p}}
\newcommand{\vk}{\mathbf{k}}
\newcommand{\vv}{\mathbf{v}}
\newcommand{\vPsi}{\boldsymbol{\Psi}}
\newcommand{\vA}{\mathbf{A}}

\newcommand{\diff}{\mathrm{d}}

\newcommand{\dd}{\delta}
\newcommand{\Dirac}{\delta_{\mathrm D}}
\newcommand{\Kronecker}{\delta^{\mathrm K}}
\newcommand{\rr}{\rho}
\newcommand{\rrb}{\bar{\rho}}
\newcommand{\ee}{\eta}

\begin{document}

\title{Emergence of Complex {\color{black} Web} Structures}

\author[1,2,3]{Francisco-Shu Kitaura}

\affiliation[1]{Cosmological Large Scale Structure Group, Instituto de Astrof\'{\i}sica de Canarias, Vía Láctea s/n, E-38205, La Laguna, Tenerife, Spain}  
\affiliation[2]{Astrophysics Department, Faculty of Physics and Mathematics, University of La Laguna, E-38206, La Laguna, Tenerife, Spain}
\affiliation[3]{Medical Technology  Division, IACTEC,  Parque  Tecnológico y Científico de las Mantecas, La Laguna, E-38320, Spain}

\emailAdd{fkitaura@iac.es}

\abstract{
Complex structures often emerge from initially homogeneous or weakly correlated states. We address the apparent tension between this ordering and entropy growth through a unified framework combining semi-microscopic phase-space dynamics, transport geometry, information theory, and coarse-grained effective modeling. The key point is that entropy depends on the level of description: a coarse-grained spatial field may become more ordered as structure forms, even while the full phase-space description becomes more complex through shell crossing, multistreaming, and the activation of velocity degrees of freedom.
Using a Lagrangian--Eulerian transport map, we show how density amplification is governed by the Jacobian of the deformation and how anisotropic collapse arises from the eigenvalues of a hierarchy of deformation tensors. Long-range interaction or information flow is encoded in the displacement field, so that nonlocality enters directly through transport. We connect this geometric description to a maximum-entropy Gaussian baseline and show how nonlinear transport and nonlocal coupling generate scale coupling, higher-order correlations, and non-Gaussianity. We then formulate a Landau--Ginzburg description in which the growth of seed anisotropies is interpreted as the activation of lower effective free-energy branches, providing a coarse-grained realization of self-organization.
Applied to generated cosmological fields, this framework indicates that the nonlocal tidal level becomes relevant already at moderate overdensity. Although cosmological structure formation is the main realization considered here, the framework is intended more broadly as a mesoscopic language for systems in which transport, anisotropy, nonlocality, and self-organization are central.
}


\maketitle
    
\section{Introduction}
\label{sec:intro}

In the quest for a unified theory of physics, several important lines of research have suggested that gravity might admit an emergent interpretation, arising from a deeper quantum, statistical, or information-theoretic substrate \citep[][]{2010RPPh...73d6901P,1995JMP....36.6377S,1995PhRvL..75.1260J,2011JHEP...04..029V,2025PhRvD.111f6001B,2025arXiv251022545B,2016PhRvD..93l4052C}. These directions are partly inspired by earlier insights into the relation between gravity, black holes, and entropy \citep[][]{1973PhRvD...7.2333B,1976PhRvD..13..191H,1977PhRvD..15.2738G,1973CMaPh..31..161B}. 

At the same time, nature repeatedly generates complex web-like patterns across a vast range of scales, from the microscale to the largest structures in the Universe. This raises a more general question: what common principles underlie the emergence of such diverse structures as {\color{black}  the cellular cytoskeleton \citep[][]{Kadzik2020,Hohmann2019,Farhadi2018,Nedelec2003,Mitchison1992},} evaporation patterns in  sessile droplets \citep[][]{Carrithers2020WhiskeyWeblike,Bennacer_Sefiane_2014}, the  human brain \citep[][]{2024PhLRv..48...47P,2014PNAS..11114247W,2015PNAS..11213455G,Vazza20,ROSELL2025121500,2025NatSR..1542309K}, the spatial distribution of trees \citep[][]{Xin2022,Abellanas_Abellanas_Pommerening_Lodares_Cuadros_2016},  the cosmic web \citep[][]{Kitaura_2021,Jasche_2019,2018MNRAS.473.1195L,10.1093/mnras/stu768,Kitaura_2012a,Nuza_2014,jaschePresentCosmicStructure2015,2010ApJ...723..364A} or even artificial neural networks \citep[][]{1998RPPh...61..353E,2015PhRvX...5b1028M,2012JPhA...45c3001B}?

Complex {\color{black} web}  structures often emerge from states that are initially homogeneous, weakly correlated, or apparently noisy. This occurs in cosmology through the formation of the cosmic web, but related questions arise much more broadly whenever transport, interaction, and nonlinear and nonlocal amplification generate coherent large-scale morphology. The central problem is therefore not simply how clustering occurs, but how ordered anisotropic structure can arise without contradicting {\color{black} entropy growth in the appropriate coarse-grained
dynamical description}.
In the present context, we use the term "complex  structure" in a precise statistical and dynamical sense: a system is complex when its morphology and evolution can no longer be captured by low-order descriptors such as the mean, the variance, or a single-stream density field, but instead require higher-order correlations, anisotropy-sensitive operators, or a more complete phase-space description.

{\color{black}
The literature on emergence emphasizes that there is no single universal
definition of complex structure \citep[see, e.g.,][]{Lichtenstein2025,Yuan2024,Green2023,Raducha2020,Baas1997}. Emergence has been described across
disciplines as the formation of coherent macroscopic patterns, properties,
or levels of organization that arise from interactions among lower-level
components and are not already evident in those components taken
separately. In network-based accounts, such patterns are generated through
interactions, topology, feedback, percolation, entrainment, modularity, and
phase transitions. Self-organization provides another
closely related viewpoint, in which large dynamical systems generate
ordered patterns without a central organizer.

Here we adopt a narrower mesoscopic definition: complex web structures are
organized spatial patterns whose description requires transport,
anisotropy, nonlocality, coarse-graining, and correlations beyond the
lowest statistical moments. This definition is complementary to broader
uses of complexity in biology, information theory, network science and
non-equilibrium thermodynamics. In particular, life is often treated as a
paradigmatic open self-organizing system, in which metabolism, regulation,
adaptation and replication maintain local order through the consumption of
external free energy. Such dissipative and biological forms of emergence
are conceptually related to the present discussion, but their specific
open-system thermodynamics and biological regulation lie outside the scope
of this work.
}

The main point of this work is that this apparent tension is resolved once one distinguishes clearly between different levels of description. A coarse-grained spatial field may become more ordered as sheets, filaments, and knots form, and the corresponding spatial entropy may decrease. At the same time, the full phase-space description becomes more complex, because shell crossing and multistreaming activate velocity degrees of freedom that are invisible in a density-only description. The same transition that generates the geometric skeleton of structure formation therefore also marks the breakdown of the simplest reduced description.

We develop this idea in a unified framework combining four ingredients. First, a semi-microscopic phase-space description makes explicit that density and velocity fields are only lower-order moments of a more complete distribution function \citep[][]{2013ApJ...762..116Y,2016JCoPh.321..644S,2016MNRAS.455.1115H,2020MNRAS.493.2765S,2021A&A...647A..66C,2024MNRAS.52710802O}. Second, a kinematic transport description based on a Lagrangian--Eulerian map explains density amplification through the Jacobian of the transformation and relates anisotropic collapse to the eigenvalues of the deformation tensor \citep[][]{Zeldovich_1970,Buchert_1994,Bouchet_1995,Catelan_1995,Bond_1996a,Monaco_1999,Hahn_2007,Forero_2009,2012MNRAS.425.2443K,Nuza_2014,2026arXiv260315834K}. Third, an information-theoretic viewpoint clarifies the distinction between reduced spatial entropy and the growing complexity of the full phase-space state \citep[][]{1957PhRv..106..620J,1957PhRv..108..171J,cover2012elements,2007spp..book.....K,1987imsm.book.....C,shannon}. Fourth, a coarse-grained effective description leads naturally to a Landau--Ginzburg-type functional  \citep[][]{landau1965collected,WAGNER19801345,SPREKELS198959,Iida:2000ha,1991JChPh..94.2176S,1994JChPh.100.2139B,1986PhRvL..57.1733A,2026PhRvF..11c4401W} in which anisotropy appears as an order parameter and web-like morphology can be interpreted as the activation of anisotropic branches (see Appendix \ref{app:anisotropy_stationarity}).

Cosmological structure formation provides the most developed realization of this logic. The primordial density field is very close to Gaussian, the linear regime admits a compact description in terms of a scalar potential, and anisotropic collapse naturally produces sheets, filaments, and knots \citep[][]{Bernardeau_2002,Peebles_1980,Angulo_2022}. Yet the purpose of the present work is broader than cosmology itself. The aim is not to claim identical microscopic dynamics across different systems, but rather to identify a common mesoscopic language in which transport, anisotropy, multistreaming, nonlocality, and higher-order correlations can be discussed in a unified way. This perspective is potentially relevant not only for cosmological large-scale structure, but also for other complex systems in which structured morphology emerges from weakly organized initial states.
The conceptual framework developed below is motivated primarily by general considerations of transport, phase-space projection, nonlocality, and coarse-graining. The numerical examples are therefore intended not as the sole basis for the argument, but as controlled illustrations of how these general mechanisms appear in a concrete cosmological realization.
This broader perspective may also be useful for systems such as the human brain, where anisotropic morphology and nonlocal organization emerge under coarse-graining even though the underlying microscopic dynamics are not gravitational.

The paper is organized as follows. In Section~\ref{sec:micro_coarse}, we distinguish the microscopic phase-space description from reduced spatial and coarse-grained descriptions, and show how spatial entropy can decrease even while phase-space complexity increases. In Section~\ref{sec:transport_web}, we introduce the transport map, Jacobian amplification, and the geometric origin of anisotropic web formation. In Section~\ref{sec:maxent_nongaussian}, we connect maximum-entropy Gaussian baselines with nonlinear mode coupling and higher-order correlations. In Section~\ref{sec:effective_theory}, we formulate a coarse-grained Landau--Ginzburg description in which anisotropy plays the role of an order parameter. Section~\ref{sec:quantitative_tests} presents numerical diagnostics on generated fields, including density--anisotropy relations, web-conditioned activation and  free-energy proxies. 
Finally, we present a discussion of the results and the conclusions (Section~\ref{sec:discussion}).

\begin{figure*}
\centering
\resizebox{\textwidth}{!}{%
\begin{tikzpicture}[
    >=Latex,
    font=\small,
    node distance=7mm and 10mm,
    box/.style={
        draw,
        rounded corners,
        align=center,
        inner sep=4pt,
        minimum height=8mm,
        fill=white
    },
    bigbox/.style={
        draw,
        rounded corners,
        inner sep=6pt,
        fill=gray!8
    },
    arr/.style={->, thick},
    darr/.style={->, thick, dashed}
]

\node[bigbox, minimum width=15.2cm, minimum height=3.6cm, anchor=north] (init) at (0,0) {};
\node[font=\bfseries] at ($(init.north)+(0,-.5)$) {Complex Structure Formation};
\node[font=\bfseries] at ($(init.north)+(0,-1.5)$) {Initial ingredients};

\node[box, minimum width=4.2cm] (system) at (-4.3,-1.5)
{system with long-range\\interaction or information flow};

\node[box, minimum width=4.2cm] (seed) at (4.3,-1.5)
{{\color{black} initial} fluctuating field\\seed of anisotropy};

\node[box, minimum width=4.3cm] (nonlocal) at (0,-2.6)
{\textbf{nonlocality}};

\draw[arr] (system) -- (nonlocal);
\draw[arr] (seed) -- (nonlocal);

\node[bigbox, minimum width=15.2cm, minimum height=3.6cm, anchor=north] (stat)
at (0,-4.3) {};
\node[font=\bfseries] at ($(stat.north)+(0,-0.5)$) {Statistical cascade};

\node[box, minimum width=3.2cm] (scales) at (-4.8,-5.9)
{coupling of\\different scales};

\node[box, minimum width=3.8cm] (hoc) at (0,-5.9)
{emergence of higher-order\\correlations};

\node[box, minimum width=3.2cm] (ng) at (4.8,-5.9)
{deviation from\\Gaussianity};

\draw[arr] (nonlocal.south) -- ($(nonlocal.south)+(0,-0.7)$) -| (scales.north);
\draw[arr] (scales) -- (hoc);
\draw[arr] (hoc) -- (ng);

\node[bigbox, minimum width=15.2cm, minimum height=4.0cm, anchor=north] (eff)
at (0,-8.6) {};
\node[font=\bfseries] at ($(eff.north)+(0,-0.5)$) {Effective coarse-grained ordering};

\node[box, minimum width=5.0cm] (growth) at (-3.9,-9.8)
{growth of seed anisotropies};

\node[box, minimum width=5.8cm] (lg) at (3.9,-9.8)
{favoured in an effective\\Landau--Ginzburg free energy};

\draw[arr] (ng.south) -- ($(ng.south)+(0,-0.7)$) -| (growth.north);
\draw[arr] (growth) -- (lg);

\node[font=\bfseries] at ($(eff.north)+(0,-2.2)$) {Self-organization};

\node[box, minimum width=5.0cm] (micro_ent) at (-3.9,-11.8)
{entropy increase in the\\ microscopic description};

\node[box, minimum width=5.8cm] (meso_ent) at (3.9,-11.8)
{entropy decrease in the\\ mesoscopic / projected description};

\draw[arr] (growth.south) -- (micro_ent.north);

\draw[arr] (micro_ent) -- (meso_ent);

\end{tikzpicture}%
}
\vspace{-0.75cm}
\caption{Summary of the conceptual chain proposed in this work. A system with long-range interaction or information flow, together with an {\color{black} initial} fluctuating field that seeds anisotropy, leads naturally to nonlocality. Nonlocality couples different scales, producing higher-order correlations and a departure from Gaussianity. In the coarse-grained description, these dynamics amplify the initial anisotropic seeds, which can be interpreted as being favoured within an effective Landau--Ginzburg free-energy framework. The lower panel highlights the self-organization aspect: entropy can increase in the semi-microscopic phase-space description while decreasing in the mesoscopic or projected description. {\color{black} The spectral properties required of this initial field are discussed in
Appendix~\ref{app:initial_spectra_web}.}}
\label{fig:framework_sketch}
\end{figure*}
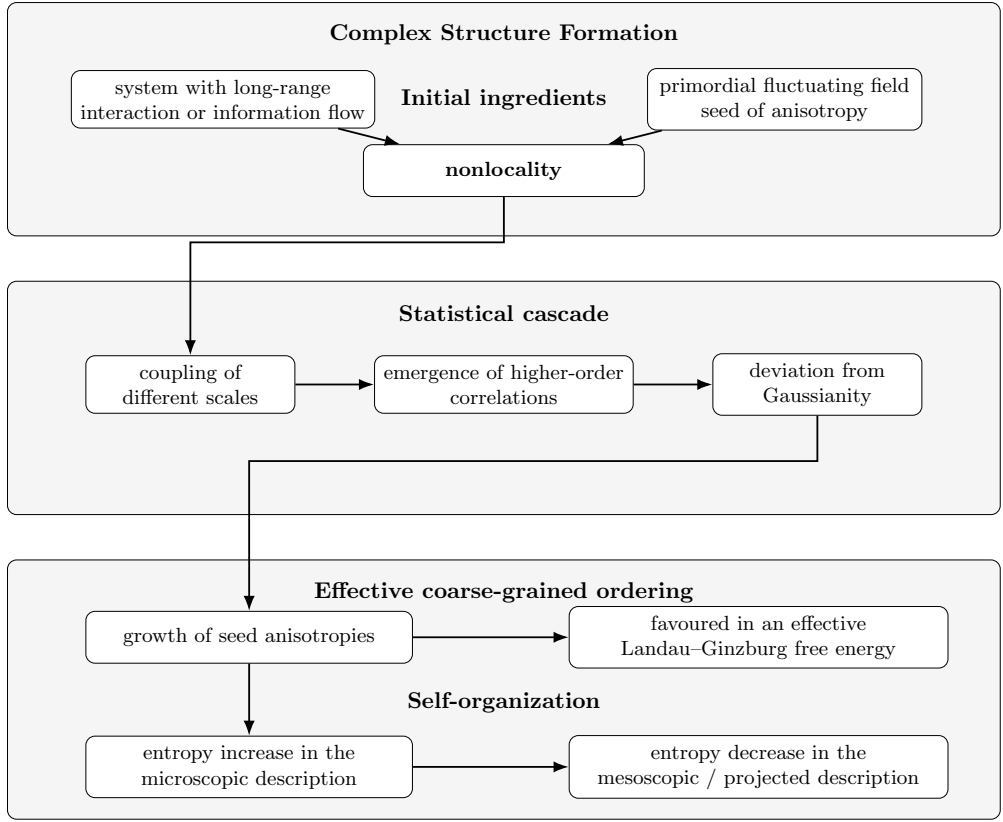

Technical derivations and fitting details are collected in the appendices.
{\color{black} In particular, Appendix~\ref{app:initial_spectra_web} discusses the
spectral properties of the initial conditions required for web-like
structure to emerge, including the Harrison--Zel'dovich scale-invariant
case and deviations from it.} Figure~\ref{fig:framework_sketch} summarizes
schematically the relationships among the main concepts discussed in this
work, starting from the initial fluctuating field that seeds anisotropy.

\section{Semi-microscopic and coarse-grained descriptions}
\label{sec:micro_coarse}

A natural semi-microscopic description of a many-particle or many-element system is formulated in phase-space through an effective one-particle distribution function \(f(t,\vx,\vv)\), such that 
$f(t,\vx,\vv)\,\diff^n x\,\diff^n v
$ gives the amount of mass, tracer weight, or coarse-grained intensity contained in the infinitesimal phase-space element around \((\vx,\vv)\). 
This should be understood as semi-microscopic rather than fully microscopic: it resolves positions and velocities in phase-space, but not the complete set of underlying constituent degrees of freedom or their exact many-body correlations\footnote{In the ideal classical limit, a system of collisionless
particles with complete phase-space information evolves reversibly and
conserves its fine-grained phase-space entropy, in accordance with Liouville's
theorem. The entropy increase discussed here therefore concerns an effectively
coarse-grained or incomplete description. In practice, some degree of
coarse-graining is unavoidable even at microscopic scales, since observations
never provide complete knowledge of the state and finite resolution cannot
resolve arbitrarily fine phase-space structure. Processes beyond the ideal closed collisionless description, including interactions, decays, annihilation or coupling to unobserved degrees of freedom, can make the reduced dynamics effectively irreversible.}. In cosmology this is the usual one-particle distribution function of collisionless matter \citep[][]{Peebles_1980,Bernardeau_2002,2010gfe..book.....M,Angulo_2022}, while in more general settings it may be interpreted more broadly as an effective state-space density over positions and velocities, or positions and flows.

The familiar density and velocity fields are reduced moments of this more complete description. In particular, the density field is obtained by integrating over velocity space, and 
the continuity equation follows as the zeroth-moment consequence of phase-space dynamics (see Appendix~\ref{app:continuity}), in the presence of source or sink terms,
\begin{equation}
\partial_t \rr+\nabla_\vx\cdot(\rr \vu)=\sigma.
\label{eq:continuity_source_main}
\end{equation} 
The crucial point is that a density-only description is incomplete: it is valid only as long as the local velocity distribution remains effectively single-stream \citep[][]{Bernardeau_2002,2011JCAP...05..015S}.

This distinction is essential for the entropy discussion. A coarse-grained spatial field may develop visible order as clustering proceeds, so that entropy measures based only on the spatial occupancy field can decrease. For example, an initially almost homogeneous density field maximizes the spatial Shannon entropy associated with a voxelized mass distribution, while the emergence of clustering redistributes probability toward many low-density cells and a smaller number of highly occupied ones, thereby reducing that particular entropy \citep[][]{cover2012elements,shannon,1994JSP....77..217P,10.1063/5.0264945} (see Appendix \ref{app:spatial_entropy}). In this projected sense, structure formation can indeed look like an entropy decrease.

However, this does not exhaust the information content of the system. Once shell crossing occurs, the local phase-space distribution is no longer described by a single velocity at each spatial location. Several streams with different velocities may coexist at the same Eulerian position. A convenient way to express this is through the entropy decomposition
\begin{equation}
S_{\rm ps}=S_x+S_{v|x},
\label{eq:entropy_chain_main}
\end{equation}
where \(S_{\rm ps}\) is the coarse-grained phase-space entropy, \(S_x\) is the coarse-grained spatial entropy, and \(S_{v|x}\) is the conditional entropy of velocities given positions. In the single-stream regime one has \(S_{v|x}=0\), so the phase-space and spatial descriptions carry equivalent information. After shell crossing, however, \(S_{v|x}>0\): the velocity field is no longer slaved to the density field, and the full phase-space state becomes less compressible even if the projected spatial field appears more ordered \citep[][]{Abel_2012,2011JCAP...05..015S,2014MNRAS.437.3442H,1982GApFD..20..111A} (see Appendix \ref{app:phase_space_entropy}).

\begin{figure}[t]
\includegraphics[width=\textwidth]{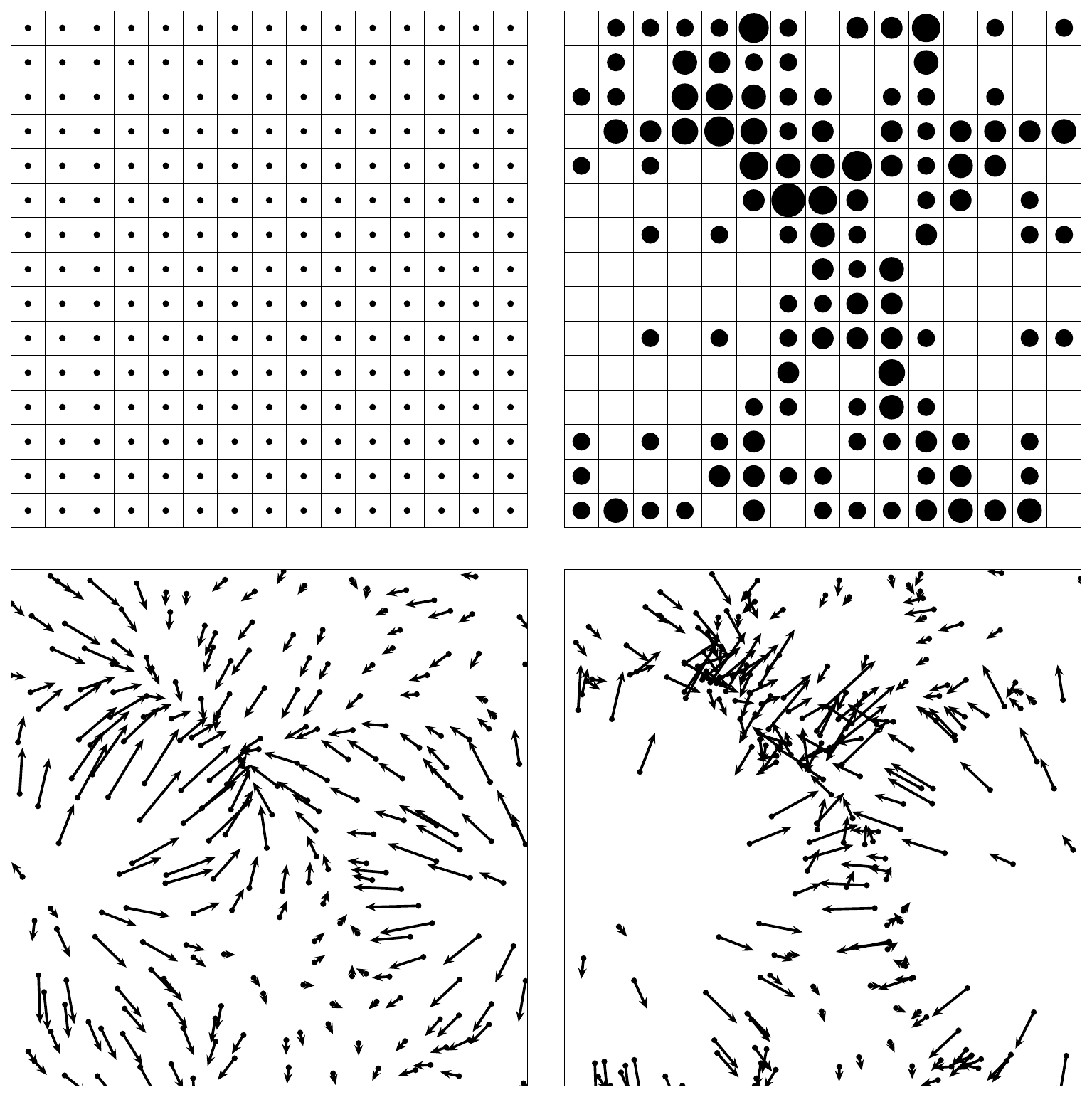}
\put(-442,435){\fcolorbox{white}{white}{\color{black}$\boxed{\rr(\vq)=\rrb\rightarrow S_\text{spatial}^\text{max}}$}}
\put(-219,435){\fcolorbox{white}{white}{\color{black}$\boxed{\rr(\vx)=\rr(\vq+D\vPsi)\rightarrow S_\text{spatial}^\text{min}}$}}
\put(-219,207){\fcolorbox{white}{white}{\color{black}$\boxed{f(t,\vx,\vv)|_\text{final}\rightarrow S_\text{phase-space}^\text{max}}$}}
\put(-442,207){\fcolorbox{white}{white}{\color{black}$\boxed{f(t,\vx,\vv)|_\text{initial}\rightarrow S_\text{phase-space}^\text{min}}$}}
\caption{Sketch illustrating the distinction between projected spatial order and microscopic phase-space complexity. \textbf{Top left}:  initially homogeneous configuration, with one tracer per cell on a regular Lagrangian mesh, representing a nearly uniform density field and high coarse-grained spatial entropy. \textbf{Top right}: the corresponding evolved spatial distribution after the Zel'dovich mapping $\mathbf{x}=\mathbf{q}+D\boldsymbol{\Psi}(\mathbf{q})$, shown as the occupancy of a fixed Eulerian grid; large empty regions and concentrated mass in fewer cells make the projected spatial distribution appear more ordered and reduce the coarse-grained spatial entropy. \textbf{Bottom row}: the same initial Gaussian field and the same set of Lagrangian tracers, now placed at random positions within each cell, shown for two stages of the evolution. \textbf{Bottom left}: microscopic phase-space representation of the initial state, with tracer positions and associated velocity vectors in the single-stream regime, where the velocity field is effectively determined by the density field.  \textbf{Bottom right}: microscopic phase-space representation after further evolution, when shell crossing has occurred and multistreaming is present. Nearby particles in configuration space can then belong to different velocity streams, making the full phase-space state more intricate even though the coarse-grained spatial map appears simpler. Black arrows indicate the associated linear-theory peculiar velocities, proportional to $\boldsymbol{\Psi}$ and rescaled for visibility. 
The figure is generated from a single realization of a Gaussian random density field on the periodic unit square with a decaying power-law spectrum, $P(k)\propto k^{n}$, with $n=-2.15$ chosen here for illustration.}
\label{fig:sketch}
\end{figure}
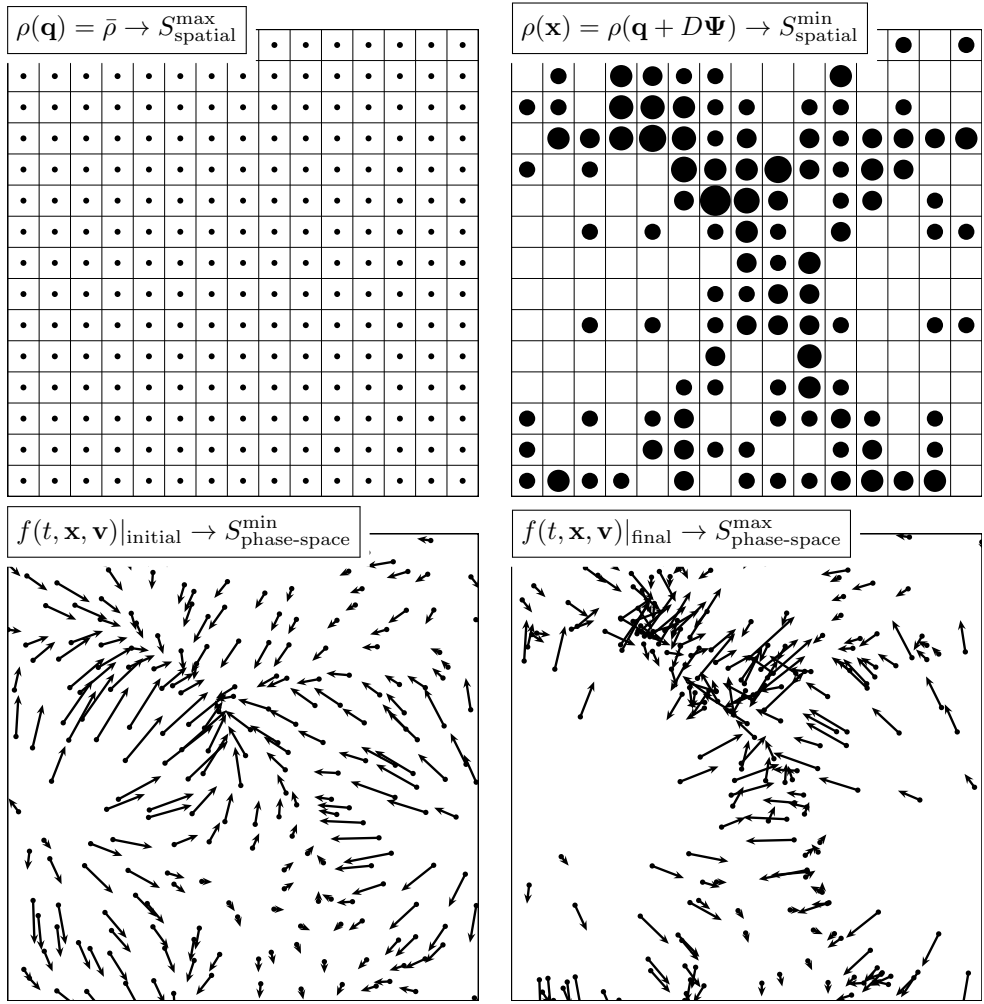

The apparent tension between entropy growth and structure formation is therefore resolved by recognizing that different entropy notions refer to different levels of description. What decreases is the entropy of a reduced spatial projection, while the full phase-space description gains complexity through multistreaming. In this sense, the emergence of coherent morphology does not contradict the growth of entropy; rather, it reflects the breakdown of reduced descriptions that discard dynamically relevant velocity information. {\color{black}
The growth of dynamical complexity has been quantified with
information-theoretic diagnostics in cosmological simulations of forming
galaxy clusters \citep[][]{Vazza17}.
}
This {\color{black} projected ordering together with increasing phase-space complexity} is
schematically shown in Figure~\ref{fig:sketch}.

This transition also has an immediate geometric meaning. The same shell crossing that activates a nontrivial conditional velocity entropy produces caustics and multistream regions, and these in turn form the structural backbone of sheets, filaments, and knots. The next step is therefore to understand how this geometric organization arises from transport itself.

\section{Transport, shell crossing and web formation}
\label{sec:transport_web}

A general way to describe structure formation is through a map between an initial configuration and an evolved one.  We therefore introduce Lagrangian coordinates \(\vq\), which label the initial state, and Eulerian coordinates \(\vx\), which describe the evolved state. The transport is encoded in the map
\begin{equation}
\vx(\vq,t)=\vq+\vPsi(\vq,t),
\label{eq:transport_map_main}
\end{equation}
where \(\vPsi\) is the displacement field \citep[][]{Zeldovich_1970,Buchert_1994,Bouchet_1995,Catelan_1995,Bernardeau_2002}.
This transport picture also admits a variational interpretation.  

\FloatBarrier

In the single-stream regime, where the map remains one-to-one, the displacement can be related to least-action and optimal-transport formulations of structure formation, in which the evolved mass distribution is connected to the initial one by a map that extremizes an action or, equivalently in suitable approximations, minimizes a quadratic transport cost \citep[][]{1989ApJ...344L..53P,2000MNRAS.313..587N,2002MNRAS.335...53B,2003MNRAS.346..501B,2003A&A...406..393M}. This variational viewpoint is most natural in the single-stream regime, before shell crossing destroys the one-to-one character of the map.

If the transported quantity is conserved, density amplification follows directly from the Jacobian of this map:
\begin{equation}
1+\dd(\vx(\vq,t),t)=J(\vq,t)^{-1},
\qquad
J(\vq,t)\equiv \det\!\left(\frac{\partial \vx}{\partial \vq}\right).
\label{eq:jacobian_density_main}
\end{equation}
In this framework, long-range interaction or information flow is encoded in the displacement field \(\vPsi(\vq,t)\), and therefore enters the density amplification through the Jacobian of the transport map.

For radially symmetric conservative interactions with conserved flux through \((n-1)\)-dimensional spheres in \(n\) spatial dimensions, the force scales as \(|\mathbf{F}(r)|\propto r^{-(n-1)}\) \citep[see, e.g.,][]{szabo1996modern,GunzburgerNicolaides1993}.  
An initially homogeneous field becomes inhomogeneous whenever the map produces nontrivial local volume deformations. More generally, source or sink terms modify this relation by an additional multiplicative factor, but the Jacobian remains the geometric core of the amplification process (see Appendix \ref{app:source_sink}).
The Jacobian also signals the onset of shell crossing. As long as \(J>0\), the map remains locally one-to-one and the flow is single-stream. When
\begin{equation}
J(\vq,t)=0,
\label{eq:shell_crossing_main}
\end{equation}
the projection onto configuration space becomes singular: several Lagrangian trajectories may then reach the same Eulerian location, and multistreaming begins \citep[][]{Angulo_2022,Abel_2012,2011JCAP...05..015S,2014MNRAS.437.3442H,1982GApFD..20..111A,Monaco_2002}. This is the precise point at which the density-only description ceases to be sufficient.


{\color{black}
A particularly transparent approximation is obtained when the displacement is potential. 
We reserve \(\phi(\mathbf q)\) for the Lagrangian displacement potential
and \(\Phi(\mathbf x)\) for the Eulerian gravitational or tidal potential.
They should not be identified without specifying the sign convention. With
the displacement convention
\begin{equation}
\vPsi(\vq,t)\simeq D(t)\nabla_{\vq}\phi(\vq),
\label{eq:potential_displacement_main}
\end{equation}
the local deformation is described by the Hessian
\begin{equation}
E_{ij}(\vq)\equiv \partial_{q_i}\partial_{q_j}\phi(\vq).
\label{eq:hessian_main}
\end{equation}
If \(\mu_a\) denote the eigenvalues of \(E_{ij}\), the Jacobian factorizes as
\begin{equation}
J(\vq,t)=\prod_{a=1}^3\left[1+D(t)\mu_a(\vq)\right].
\label{eq:jacobian_eigen_main}
\end{equation}
Shell crossing occurs when \(1+D\mu_a=0\), which requires
\(\mu_a<0\). This convention should be distinguished from the common
tidal-web convention, in which the potential is defined with the opposite
sign, \(\nabla^2\Phi=\delta\), and collapse is associated with positive
tidal eigenvalues \(\lambda_a\). In that notation one has
\(\lambda_a=-\mu_a\), so that the same condition becomes
\[
J=\prod_a(1-D\lambda_a),
\qquad
1-D\lambda_a=0.
\]
The commonly used rule \(\lambda_a>\lambda_{\rm th}>0\) is therefore a
positive-threshold web-classification criterion, not a contradiction of
the \(1+D\mu_a=0\) shell-crossing condition.
The same eigenvalue structure immediately explains the sequence of
anisotropic collapse.
}

Collapse along one principal direction produces a sheet, collapse along two directions produces a filament, and collapse along three directions in three dimensions produces a knot \citep[][]{Zeldovich_1970,Bond_1996a,Hahn_2007,Forero_2009}. In this way, the eigenvalue structure of the Hessian provides a natural geometric origin for web-like morphology.

In cosmology, this is the familiar Zel'dovich picture \citep[][]{Zeldovich_1970,Bernardeau_2002,Peebles_1980}. The key point for the present work is broader: web formation does not require isotropy to be broken by hand. Exact local isotropy would correspond to degenerate eigenvalues and therefore represents a special rather than a generic case. Starting from a noisy weakly structured field, anisotropic collapse is statistically natural, and web-like structures emerge as the generic imprint of the deformation tensor \citep[][]{Bond_1996a,Hahn_2007,Forero_2009}.

This geometric viewpoint also clarifies the meaning of shell crossing. For collisionless matter, it should not be interpreted as the literal intersection of trajectories in full phase-space, where the fine-grained sheet remains continuous. Rather, shell crossing is a singularity of the projection from phase-space onto configuration space \citep[][]{Abel_2012,2011JCAP...05..015S,2013ApJ...762..116Y,2016JCoPh.321..644S,2016MNRAS.455.1115H,2020MNRAS.493.2765S,2021A&A...647A..66C,2024MNRAS.52710802O}. The cosmic web may therefore be interpreted as the observable imprint of a folded multistream phase-space organization.

More generally, the displacement may be decomposed into irrotational and
solenoidal parts,
{\color{black}\begin{equation}
\vPsi(\vq,t)
=
D(t)\nabla_{\vq}\phi(\vq)
+
\nabla_{\vq}\times \vA(\vq,t),
\end{equation}
}
where \(\phi\) is the Lagrangian displacement potential and \(\vA\) is a
vector potential whose curl generates the divergence-free part of the
displacement.

In a Helmholtz sense, this corresponds to separating the transport into a compressive component and a rotational or shear-like component. The solenoidal term does not invalidate the basic picture developed here. Rather, it enriches it by adding rotational deformations, modifying the local Jacobian beyond the pure Hessian form, and allowing for more intricate phase-space folding and velocity structure. However, it is not required in order to obtain web-like morphology: anisotropic collapse already arises generically from the potential part alone. In that sense, the potential approximation should be regarded as a minimal model of web formation, while the solenoidal contribution captures additional complexity beyond that minimal picture. Nonetheless, the solenoidal component may become important in some cases and can be modeled either explicitly \citep[][]{Catelan_1995,2026arXiv260313106K} or implicitly through subsequent coordinate transformations \citep[][]{Kitaura_2024} (see also \citep[][]{2011JCAP...04..032M,2012JCAP...01..019P,Rampf_2012,2017PhRvD..95f3527C,2005A&A...438..443B}).

\section{Maximum entropy, nonlocality, and higher-order structure}
\label{sec:maxent_nongaussian}

The previous section explained how transport and anisotropic deformation generate web-like morphology geometrically. We now turn to the statistical side of the problem. The relevant question is not whether the field is structured, but relative to what baseline that structure should be measured.

A natural inference principle is provided by the maximum-entropy approach of Jaynes: when only incomplete information is available, the least biased probability distribution is the one that maximizes entropy subject to the known constraints \citep[][]{1957PhRv..106..620J,1957PhRv..108..171J}. Applied to field configurations, this implies that if only the mean and two-point correlation function are fixed, the maximum-entropy ensemble is Gaussian,
\begin{equation}
P[\varphi]\propto \exp\!\left[-\mathcal S[\varphi]\right],
\label{eq:maxent_gaussian_main}
\end{equation}
with a quadratic statistical action \(\mathcal S[\varphi]\). In this sense, a Gaussian random field is the least biased ensemble consistent with a prescribed covariance structure \citep[][]{cover2012elements,adler1981geometry} (see Appendix \ref{app:maxent}).

This Gaussian baseline is especially natural in cosmology, where the {\color{black} initial} field is very close to Gaussian and linear evolution preserves that property to good approximation \citep[][]{Bardeen86,Peebles_1980,Bernardeau_2002}. More generally, it provides a low-order reference state against which nonlinear organization can be assessed. Its limitation is equally clear: a Gaussian field contains no connected correlations beyond second order and therefore cannot encode the higher-order spatial dependence associated with complex morphology.

The departure from Gaussianity is driven by nonlocal interaction and nonlinear transport. A generic transport potential may be written as a nonlocal response of the field,
\begin{equation}
\phi(\vx)=\int \diff^n y\,K(\vx-\vy)\,\varphi(\vy),
\label{eq:kernel_main}
\end{equation}
with the Poisson relation of cosmology providing an important special case \citep[][]{Bernardeau_2002,Peebles_1980}. Once such a nonlocal interaction is inserted into a nonlinear transport law or into perturbative evolution equations, Fourier modes cease to evolve independently and become coupled. The evolved field is then no longer described by the power spectrum alone, but requires higher connected moments.

This is the statistical meaning of mode coupling. Even if the initial field is Gaussian, the nonlinear Jacobian map and the nonlocal transport generate higher-order correlations. In Fourier space, already the second-order contribution has the schematic form
\begin{equation}
\dd^{(2)}(\vk)\sim
\int \frac{\diff^n p}{(2\pi)^n}\,
F_2(\vp,\vk-\vp)\,
\varphi(\vp)\varphi(\vk-\vp),
\label{eq:mode_coupling_main}
\end{equation}
and higher orders generate analogous couplings among three or more modes \citep[][]{Bernardeau_2002} (see Appendix \ref{app:ept}
). The statistical consequence is that skewness, kurtosis, bispectra, trispectra, and higher connected correlators become nonzero \citep[][]{Groth77,Baumgart91,Frieman94,Matarrese97,Verde98,2021JCAP...01..015G,2021JCAP...07..008G,2016JCAP...06..052B,2001ApJ...553...14V,2021PhRvD.103j3518S,2012MNRAS.420.2737K,Kitaura_2015}.

From this viewpoint, the transition from a Gaussian to a non-Gaussian field is not merely a change in the shape of a one-point PDF. It signals the activation of higher-order dependencies across space. Higher-order correlations are therefore not only useful descriptors of structure, but the statistical imprint of nonlinear and nonlocal dynamics. In this sense, the emergence of complex morphology is tied simultaneously to geometric anisotropic collapse and to the breakdown of a Gaussian two-point statistical description.

\section{Coarse-grained effective theory of anisotropy}
\label{sec:effective_theory}

The previous sections motivate a mesoscopic description in which the relevant variables are no longer the microscopic degrees of freedom themselves, but coarse-grained fields and the local invariants that can be constructed from them. In this regime, the natural language is that of effective free-energy functionals \citep[][]{landau1965collected} (see Appendix \ref{app:LG_motivation}).

The conceptual step is that coarse-graining changes the natural variational language. At the semi-microscopic level one describes the system in terms of entropy over accessible states, but after integrating out unresolved degrees of freedom the remaining mesoscopic fields are governed by an effective functional that combines energetic and entropic contributions. In equilibrium language this is naturally written as an effective free energy (relating internal energy $U$,   temperature $T$ and entropy $S$),
\begin{equation}
F=U-TS,
\label{eq:free_energy_main}
\end{equation}
or, more generally, as a coarse-grained functional summarizing the competition between ordering tendencies and unresolved microscopic complexity. 

In this sense, the relevant variational principle changes from maximizing entropy at the microscopic level to minimizing an effective free-energy functional at the mesoscopic level.

A natural coarse-grained description is then of Landau--Ginzburg type. Rather than following all microscopic degrees of freedom individually, one writes the most general local functional consistent with the relevant symmetries and scales, constructed from density-like amplitudes, gradients, and anisotropy-sensitive operators \citep[][]{landau1965collected,Desjacques_2018,McDonald_2009,Schmittfull_2019,Pellejero_2020,Kitaura_2022,Coloma_2024,Sinigaglia_2024}. In the present context, a minimal functional depends on the density contrast \(\dd\) together with a hierarchy of Hessian-like tensors \(E^{(\ell)}_{ij}\), which encode anisotropy across different effective levels of nonlocality \citep[][]{2026arXiv260315834K}.

{\color{black}
To make this explicit, we distinguish between the total Hessian amplitude
and the traceless anisotropy invariant. For each hierarchy level \(\ell\),
define
\begin{equation}
\tilde Q^{(\ell)}(\vx)
\equiv
E^{(\ell)}_{ij}E^{(\ell)}_{ji}
=
\sum_i \left(\mu_i^{(\ell)}\right)^2
\ge 0 ,
\label{eq:Qtilde_main}
\end{equation}
where \(\mu_i^{(\ell)}\) are the eigenvalues of \(E^{(\ell)}_{ij}\). This
quantity measures the total amplitude of the Hessian, including its
isotropic trace part. The pure anisotropy invariant is instead the
traceless contraction
\begin{equation}
Q^{(\ell)}(\vx)
\equiv
\left(E^{(\ell)}_{ij}
-\frac{1}{3}\delta^{\rm K}_{ij}E^{(\ell)}_{kk}\right)
\left(E^{(\ell)}_{ij}
-\frac{1}{3}\delta^{\rm K}_{ij}E^{(\ell)}_{kk}\right).
\label{eq:Q_main}
\end{equation}
Equivalently, in terms of eigenvalues,
\begin{equation}
Q^{(\ell)}
=
\sum_i
\left(\mu_i^{(\ell)}-\bar\mu^{(\ell)}\right)^2
=
\tilde Q^{(\ell)}
-
3\left(\bar\mu^{(\ell)}\right)^2,
\qquad
\bar\mu^{(\ell)}
=
\frac{1}{3}\sum_i\mu_i^{(\ell)}.
\end{equation}
Thus \(Q^{(\ell)}\) vanishes for locally isotropic deformations and grows
with eigenvalue splitting. In what follows \(Q^{(\ell)}\) is used as the
anisotropy order parameter entering the effective Landau--Ginzburg branch,
while \(\tilde Q^{(\ell)}\) denotes the corresponding trace-containing
Hessian amplitude.
}

An interesting connection with standard cosmological bias expansions is
that the lowest-level anisotropy invariant \(Q^{(1)}\) is directly related
to the familiar long-range tidal operator. For
\(E^{(1)}_{ij}\propto \partial_i\partial_j\nabla^{-2}\delta\), the total
Hessian amplitude is
\[
\tilde Q^{(1)}=E^{(1)}_{ij}E^{(1)}_{ji},
\]
while the usual traceless tidal tensor is
\[
s_{ij}
=
\left(\partial_i\partial_j\nabla^{-2}
-\frac{1}{3}\delta^{\rm K}_{ij}\right)\delta.
\]
Hence
\[
s^2=s_{ij}s_{ij}
=
\tilde Q^{(1)}-\frac{1}{3}\delta^2
=
Q^{(1)},
\]
up to normalization conventions. In this sense, the first anisotropy level
used here coincides with the standard nonlocal tidal operator of
cosmological bias theory. The activation of the \(Q^{(1)}\) branch can
therefore be viewed as the effective-theory counterpart of the tidal
response of tracers to their large-scale environment.

\begin{figure}
\includegraphics[width=\textwidth]{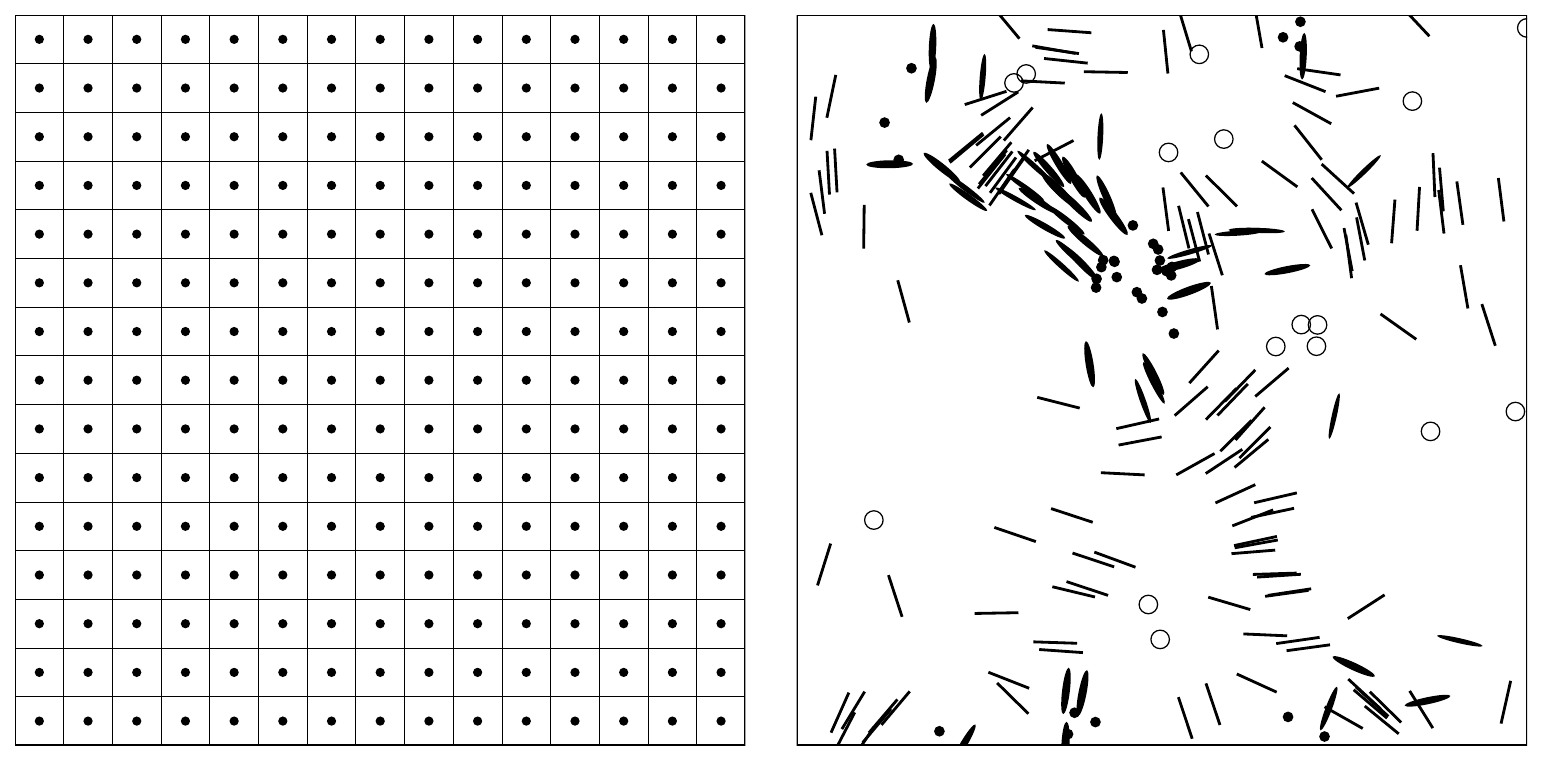}
\put(-442,215){\fcolorbox{white}{white}{\color{black}$\boxed{\rr(\vq)=\rrb,\; Q_\star=0,\; f_{\rm iso}<f_{\rm aniso}}$}}
\put(-219,215){\fcolorbox{white}{white}{\color{black}$\boxed{\rr(\vx)=\rr(\vq+D\vPsi),\; Q_\star>0,\; f_{\rm aniso}<f_{\rm iso}}$}}
\caption{Sketch illustrating a two-dimensional proxy for anisotropic web classification and its Landau--Ginzburg interpretation. \textbf{Left}: homogeneous reference state on a regular Lagrangian mesh, with isotropic coarse-grained minimum \(Q=0\). \textbf{Right}: evolved state after the mapping \(\vx=\vq+D\vPsi\), where anisotropic structures are activated and an anisotropic branch with \(Q>0\) becomes preferred. In two dimensions the Hessian \(\partial_i\partial_j\Phi\) has only two eigenvalues, so the standard three-dimensional cosmic-web rule cannot be applied unchanged. We retain the void signature --empty circles-- (both eigenvalues negative) and identify sheets with saddle points --thin lines-- (\(\lambda_1>0\), \(\lambda_2<0\)), which are the natural two-dimensional analogue of collapse along one preferred direction. Regions with both eigenvalues positive are then split schematically by anisotropy: nearly isotropic positive curvature is represented as a knot --filled circles--, whereas strongly anisotropic positive curvature is represented as a filament-like elongated structure --elongated ellipses--. The classification criterion for this split is \(\lambda_1/\lambda_2 \geq R\) for $\lambda_1,\lambda_2>0$, with \(R=2.8\). This should be understood as a two-dimensional morphological proxy rather than as the full three-dimensional eigenvalue classification. The right panel is local in its visual representation but nonlocal in its informational content: each glyph is attached to a local tracer position, yet its web classification is derived from Hessian-like operators that encode the surrounding environment and therefore compress nonlocal structural information into a local morphological label. At each tracer we diagonalise the interpolated tidal Hessian \(H_{ij}=\partial_i\partial_j\Phi\), with eigenvalues \(\lambda_1\ge\lambda_2\) and corresponding orthonormal eigenvectors \(\mathbf{v}_1\), \(\mathbf{v}_2\). {Filaments} are drawn as filled ellipses whose major axis is aligned with \(\mathbf{v}_2\), i.e.\ the direction of the smaller eigenvalue, so that the glyph is elongated along the softer principal curvature of \(\Phi\). {Sheets} are drawn as line segments oriented along \(\mathbf{v}_1\), the eigenvector of the larger eigenvalue; for the saddle configuration \((\lambda_1>0,\lambda_2<0)\) this is the direction associated with the positive eigenvalue. The ellipse rotation angle in the plot is measured counter-clockwise from the \(x\)-axis, \(\mathrm{atan2}(v_{2y},v_{2x})\). In the Landau--Ginzburg picture, the left panel represents the isotropic minimum of the local coarse-grained functional \(f_{\rm iso}\), while the right panel represents the activation of an anisotropic minimum with lower effective free-energy cost \(f_{\rm aniso}\).}
\label{fig:sketch2}
\end{figure}

A minimal local coarse-grained functional may then be written schematically as
\begin{equation}
f_{\rm loc}(\dd,\{Q^{(\ell)}\})
=
f_\dd(\dd)
+
\sum_\ell
\left[
\alpha_\ell(\dd)\,Q^{(\ell)}
+
\beta_\ell \bigl(Q^{(\ell)}\bigr)^2
\right]
+\cdots,
\label{eq:floc_main}
\end{equation}
with \(\beta_\ell>0\) for stability. The coefficient \(\alpha_\ell(\dd)\) acts as an effective density-dependent control parameter for anisotropy. Minimizing the local free-energy density with respect to \(Q^{(\ell)}\) gives the stationary branch
\begin{equation}
Q^{(\ell)}_\star(\dd)
=
\max\!\left[
0,\,
-\frac{\alpha_\ell(\dd)}{2\beta_\ell}
\right].
\label{eq:Qstar_main}
\end{equation}
This has the standard Landau interpretation: for \(\alpha_\ell(\dd)>0\), the isotropic branch \(Q^{(\ell)}=0\) is preferred; when \(\alpha_\ell(\dd)\) changes sign and becomes negative, an anisotropic branch with \(Q^{(\ell)}_\star>0\) becomes energetically favored, while \(\beta_\ell>0\) ensures stability \citep[][]{landau1965collected}.
A schematic representation is presented in Figure \ref{fig:sketch2}.

This effective sign change is physically motivated by the growth dynamics itself: in the separable growing-mode regime, the same growth factor amplifies the pre-existing eigenvalue splitting of the deformation tensor, so anisotropic collapse is the generic route to structure formation; see also the classic analysis of Gaussian initial deformation fields by {\color{black} Doroshkevich 1970} \cite[][]{1970Afz.....6..581D} (Appendix~\ref{app:anisotropic_branch_growth}).

The significance of this construction is not that all structure-forming systems literally thermalize to a common equilibrium free energy. Rather, the Landau--Ginzburg functional provides an effective operator expansion for morphology after coarse-graining. The local powers of \(\dd\) control the amplitude dependence, gradient terms penalize sharp spatial variations, and the quadratic and higher-order invariants of \(E^{(\ell)}_{ij}\) encode anisotropy and nonlocal environmental dependence. In this way, anisotropic web formation can be reinterpreted as the activation of coarse-grained branches that lower the effective free-energy cost.

This perspective is closely related to nonlocal tracer expansions in cosmology, where the response of halos or galaxies depends not only on the local density but also on tidal and curvature operators \citep[][]{Fry93,McDonald_2009,Kitaura_2014,Desjacques_2018,Chan_2012,2018MNRAS.476.3631P,2018JCAP...01..053M,Kitaura_2022,Coloma_2024,Sinigaglia_2024} and effective field theories in cosmology \citep[see, e.g.,][]{Carrasco_2012,Baumann_2012,Pajer_2013,2014JCAP...05..022P,2014JCAP...03..006M,2015JCAP...10..039A,2015PhRvD..92l3007B,2016JCAP...03..017B,2016JCAP...05..027F}.  The present formulation generalizes that logic into a hierarchy of scale-dependent anisotropy operators. It is therefore best understood as a coarse-grained effective theory of structure formation, capable of organizing density, anisotropy, and web-like morphology within a single variational framework.

\section{Quantitative tests on generated fields}
\label{sec:quantitative_tests}

The effective framework developed above leads to several qualitative expectations. First, anisotropy should become activated above a characteristic density scale. Second, this activation should depend on web environment, so that different morphological classes populate different anisotropic branches. Third, the onset of anisotropy should admit an interpretation in terms of an effective coarse-grained free-energy functional. These expectations follow from the theoretical framework itself; the role of the numerical analysis in this section is to illustrate them in a controlled cosmological realization and to assess to what extent the simplest truncations capture the observed trends.

To this end, we consider fields generated with  ridged Augmented Lagrangian Perturbation Theory (RLPT) \citep[][]{Kitaura_2013,Kitaura_2024,2026arXiv260313106K}. The simulation volume is a cube of side length \(L=100\,h^{-1}\mathrm{Mpc}\), sampled with \(400^3\) particles, corresponding to a spatial resolution \(\mathrm{d}L=0.25\,h^{-1}\mathrm{Mpc}\), for a cosmology consistent with Planck 2018 \citep{Planck2018}. Although the numerical example is cosmological, the logic of the diagnostics is more general: the relevant ingredients are a coarse-grained density field, anisotropy-sensitive operators, and a hierarchy of web environments.

\begin{figure}
    \vspace{-1.2cm}
    \hspace{-0.3cm}
    \begin{tabular}{cc}
\subfigure{\includegraphics[width=0.5\textwidth]{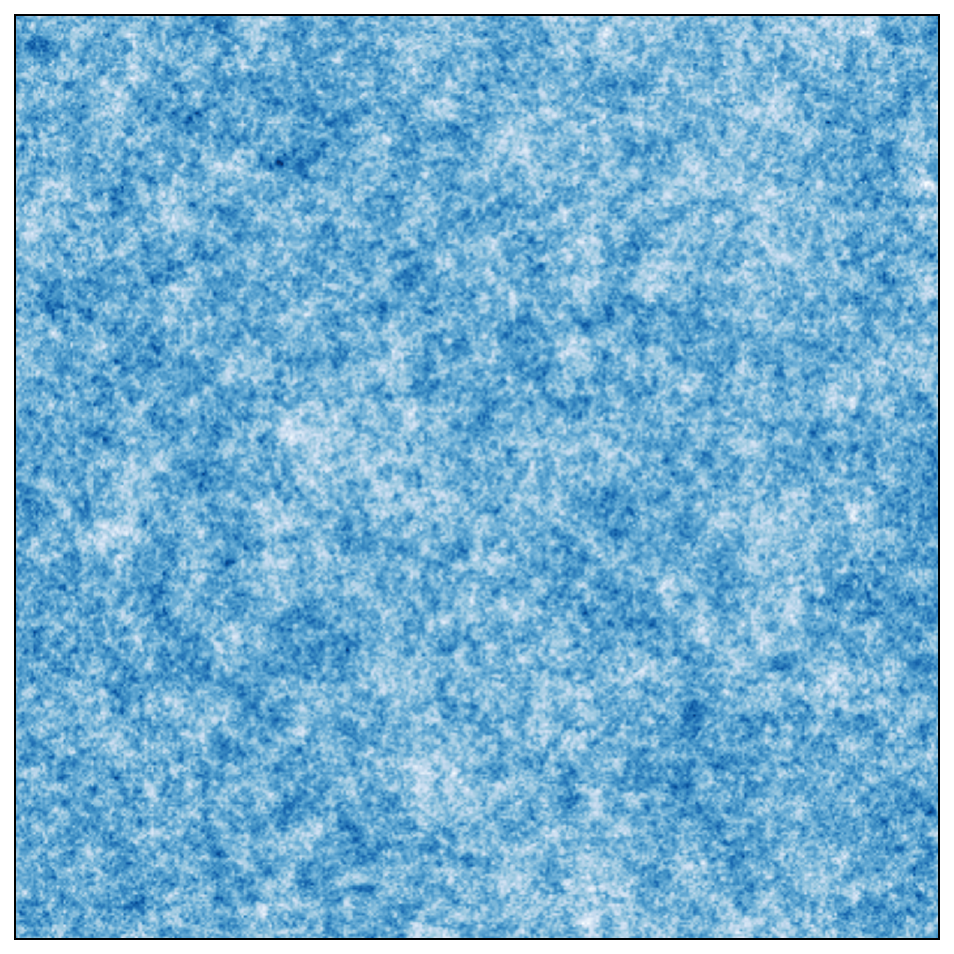}}
\put(-217,206){\fcolorbox{white}{white}{\color{black}$\dd(\vq)$}}
\hspace{-0.4cm}
&\subfigure{\includegraphics[width=0.5\textwidth]{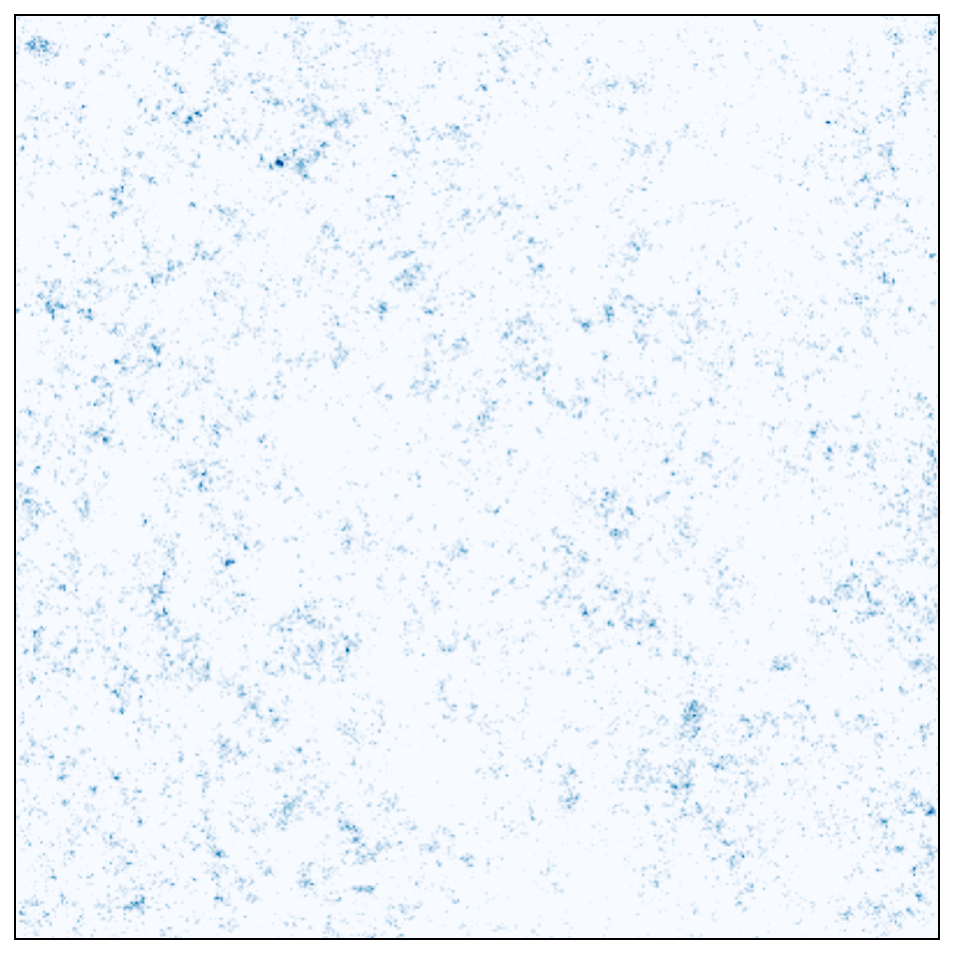}}
\put(-217,206){\fcolorbox{white}{white}{\color{black}$\exp(\dd(\vq)-\langle\dd(\vq)^2\rangle/2)-1$}}
\vspace{-0.4cm}
\\
\subfigure{\includegraphics[width=0.5\textwidth]{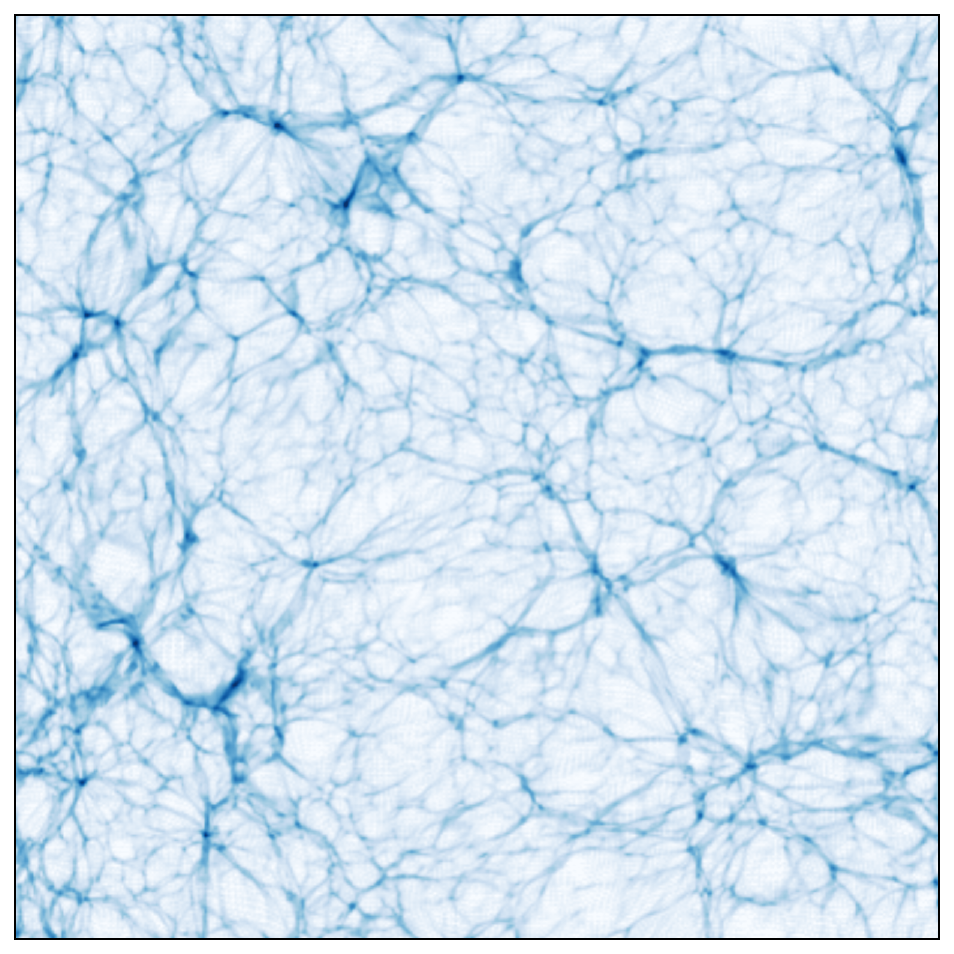}}
\put(-217,206){\fcolorbox{white}{white}{\color{black}$\dd(\vx)$}}\hspace{-0.4cm}
&
\subfigure{\includegraphics[width=0.5\textwidth]{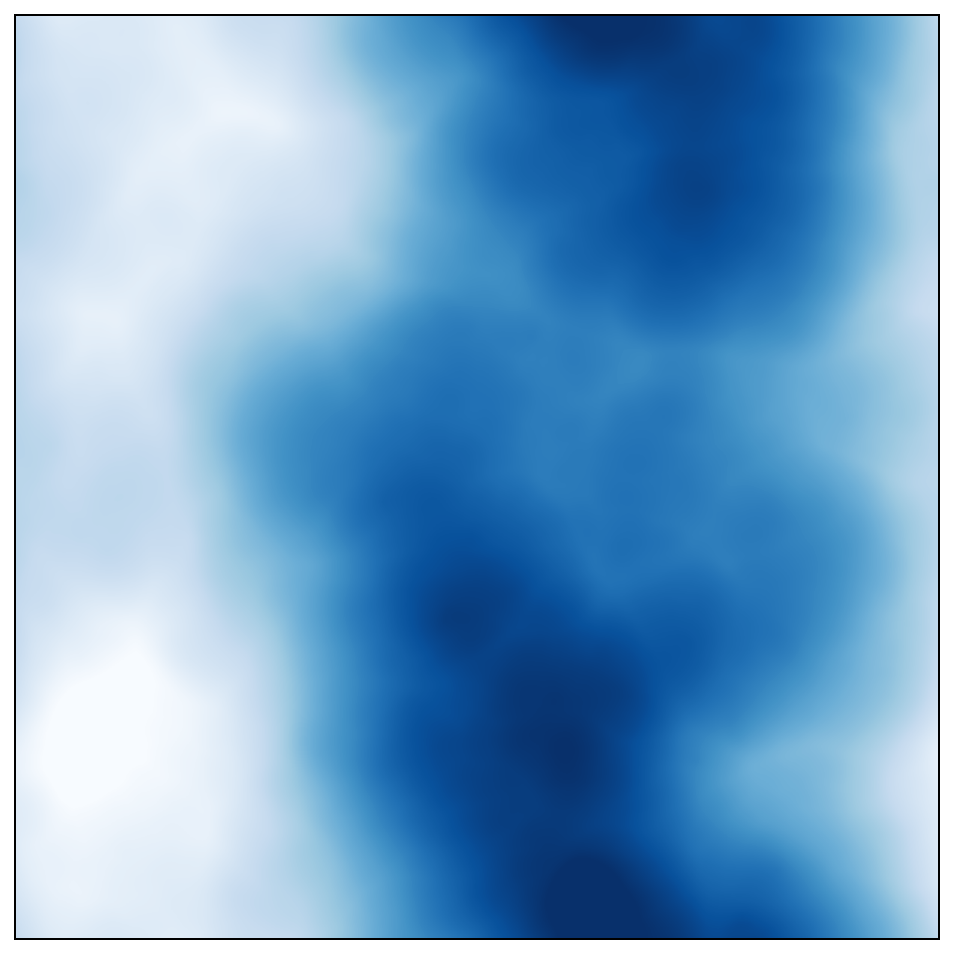}}
\put(-217,206){\fcolorbox{white}{white}{\color{black}$\Phi(\vx)=\nabla^{-2}\dd(\vx)$}}
\vspace{-0.cm}
\end{tabular}
\caption{{\bf Top left}: Two-dimensional slice through a three-dimensional cubic mesh containing a Gaussian random realization with a power spectrum that decays as a power law towards small scales. The box side length is \(100\,h^{-1}{\rm Mpc}\) and the cell size is \({\rm d}L \simeq 0.25\,h^{-1}{\rm Mpc}\), with \(h=0.677\). {\bf Top right}: Corresponding lognormal field constructed from the Gaussian realization shown in the top-left panel \citep[][]{Coles_1991,2012MNRAS.425.2443K}. 
{\bf Bottom left}: Field evolved dynamically with ridged augmented Lagrangian perturbation theory to redshift \(z=1\). {\bf Bottom right}: Potential associated with the evolved field in the bottom-left panel, obtained by applying the inverse Laplacian operator. The top-right panel shows that a local nonlinear mapping  can alter the one-point distribution of the field, but does not reproduce the filamentary connectivity and anisotropic morphology generated by nonlocal transformations or  transport.}
\label{fig:cosmic-web1}
\end{figure}

\begin{figure}
    \vspace{-1.2cm}
    \hspace{-0.3cm}
    \begin{tabular}{cc}
\subfigure{\includegraphics[width=0.5\textwidth]{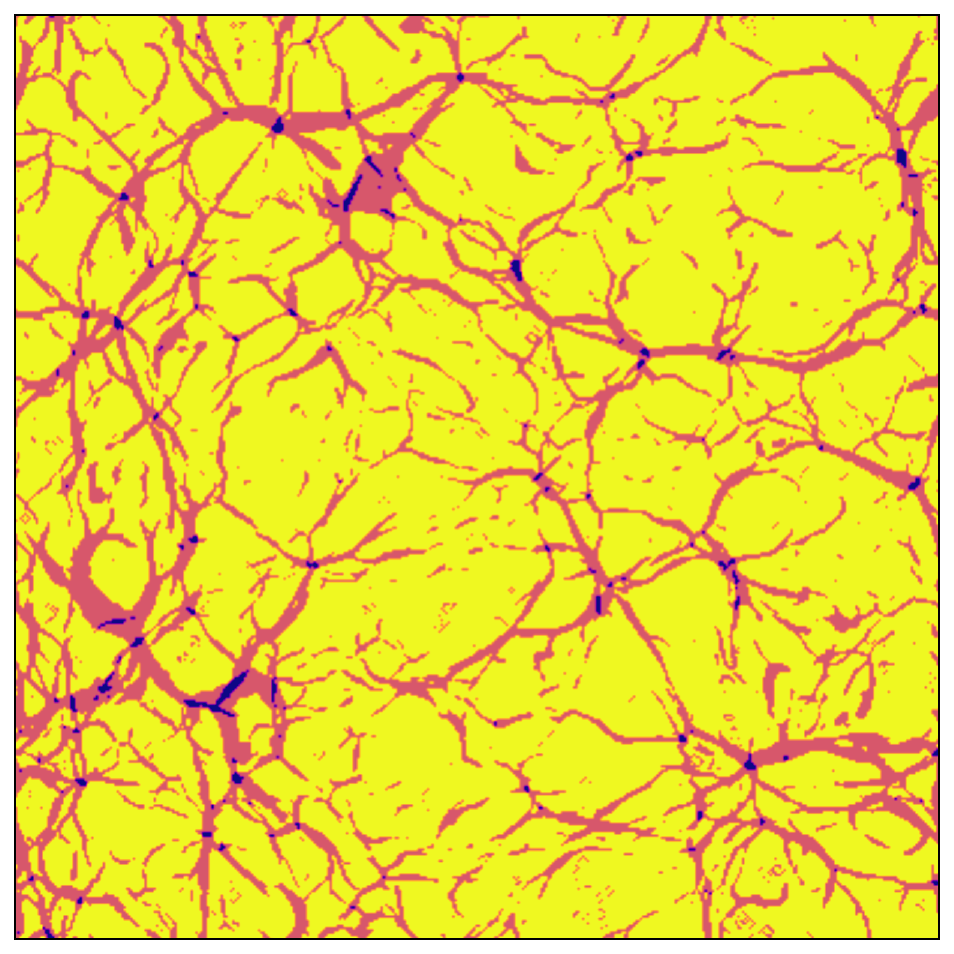}}
\put(-217,206){\fcolorbox{white}{white}{\color{black}$E^{(1)}_{ij}=\nabla_{ij}\nabla^{-2}\dd(\vx)$}}
\hspace{-0.4cm}
&
\subfigure{\includegraphics[width=0.5\textwidth]{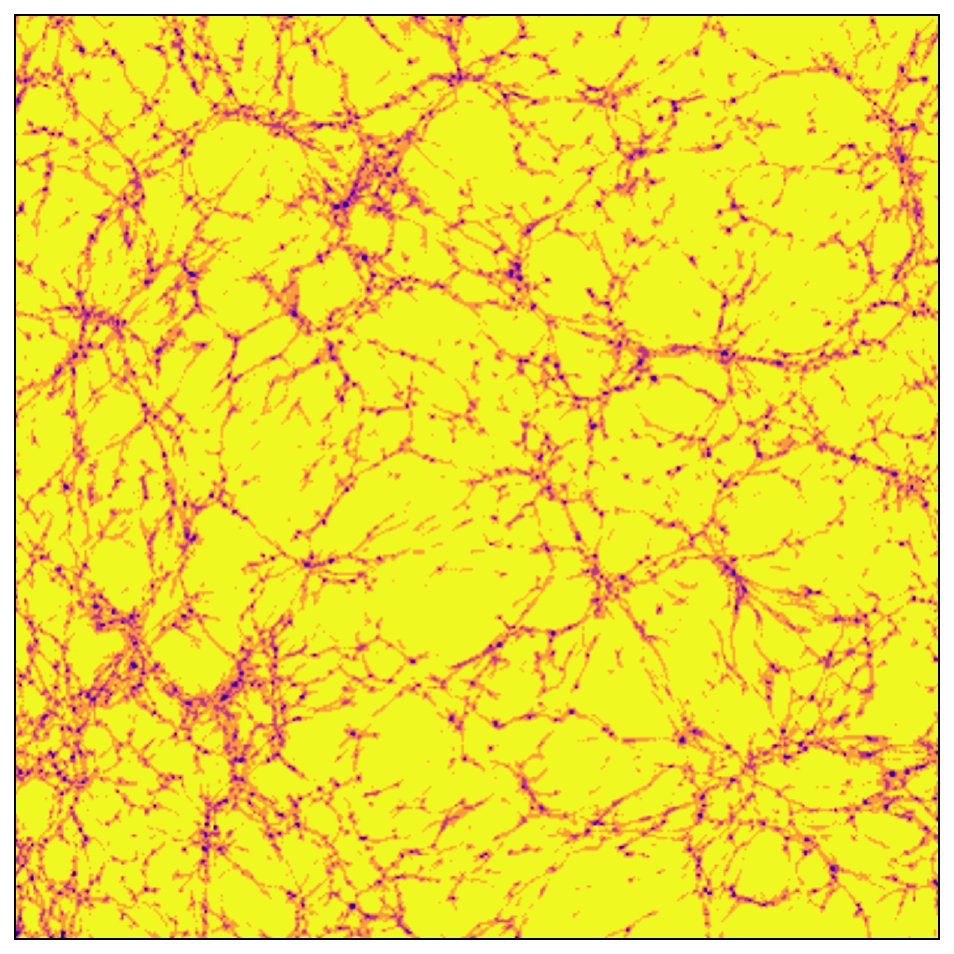}}
\put(-217,206){\fcolorbox{white}{white}{\color{black}$E^{(2)}_{ij}=\nabla_{ij}\dd(\vx)$}}\vspace{-0.4cm}
\\
\subfigure{\includegraphics[width=0.5\textwidth]{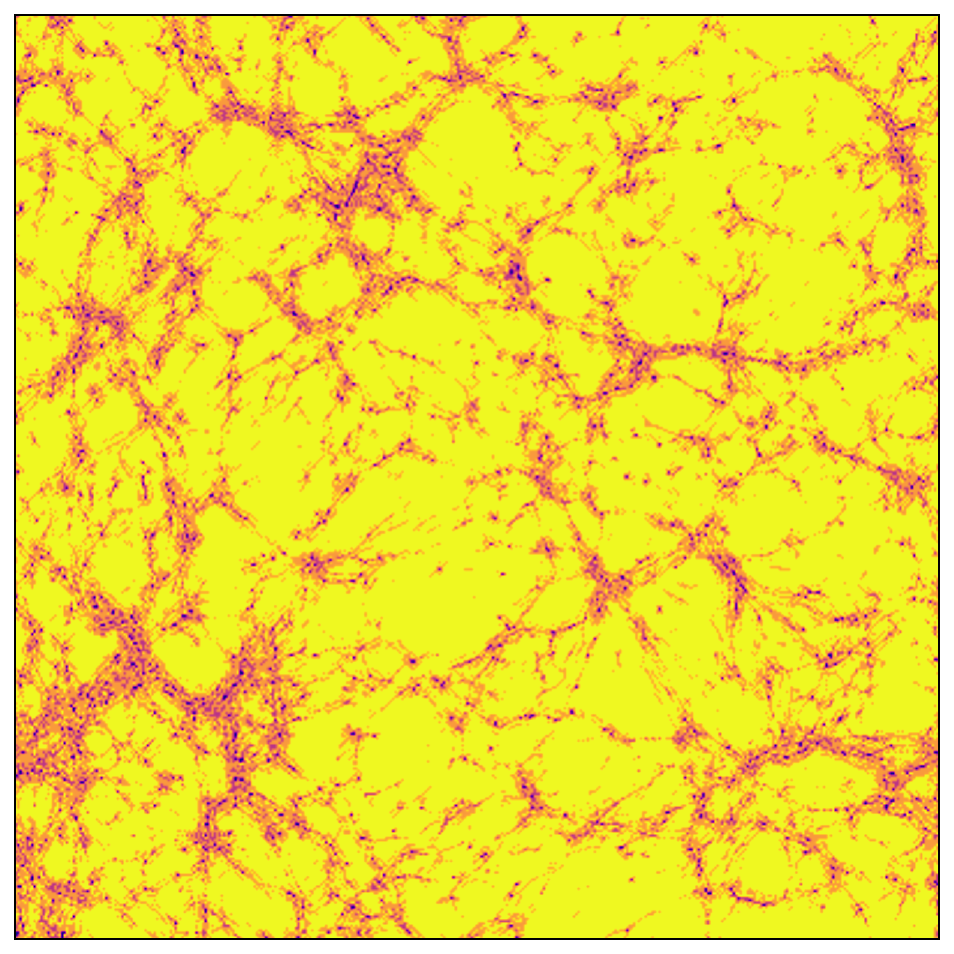}}
\put(-217,206){\fcolorbox{white}{white}{\color{black}$E^{(3)}_{ij}=\nabla_{ij}\nabla^{2}\dd(\vx)$}}\hspace{-0.4cm}
&
\subfigure{\includegraphics[width=0.5\textwidth]{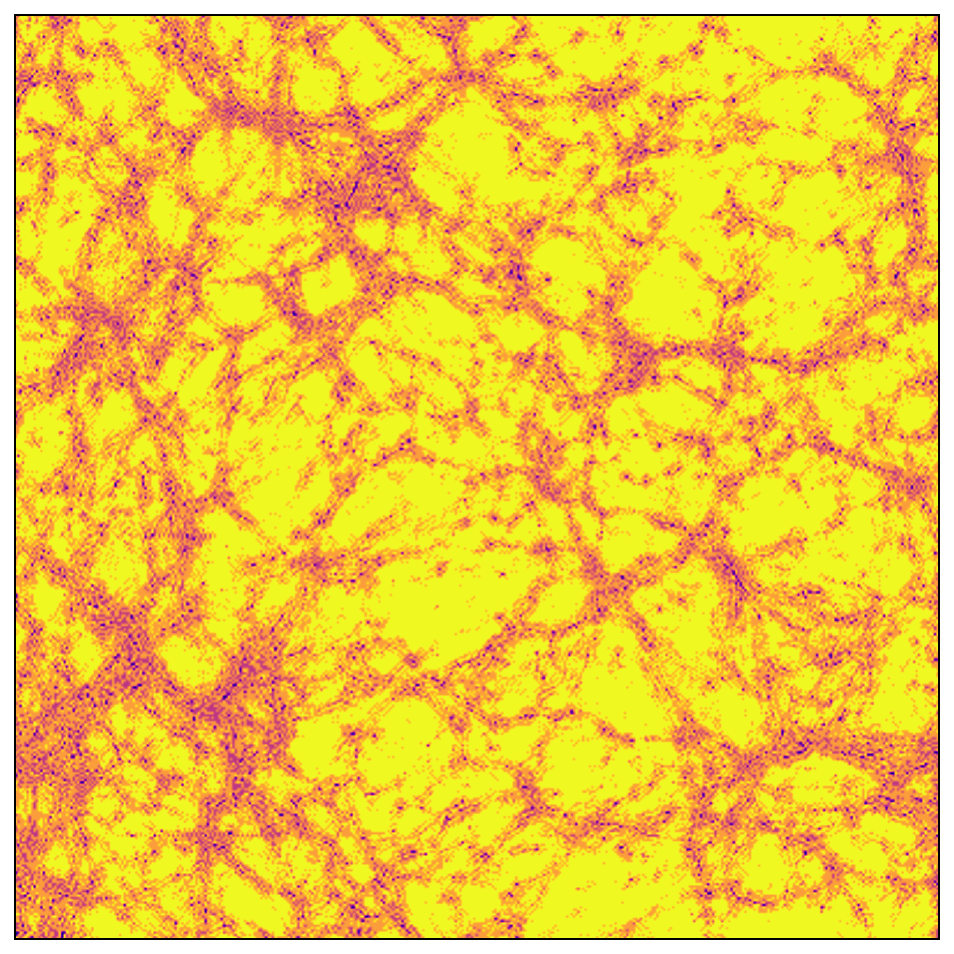}}
\put(-217,206){\fcolorbox{white}{white}{\color{black}$E^{(4)}_{ij}=\nabla_{ij}\nabla^{4}\dd(\vx)$}}\vspace{-0.cm}
\end{tabular}
\caption{Web contrasts derived from the Hessian-like tensors corresponding to the spectral hierarchy \(\nabla_{ij}\nabla^{-2}\delta(\vx)\), \(\nabla_{ij}\delta(\vx)\), \(\nabla_{ij}\nabla^{2}\delta(\vx)\), and \(\nabla_{ij}\nabla^{4}\delta(\vx)\). The displayed values are discrete web labels, with \(4=\) knots, \(3=\) filaments, \(2=\) sheets, and \(1=\) voids. The sequence shows how the inferred morphology changes from large-scale tidal structure to progressively more local curvature-sensitive web contrasts: \(\nabla_{ij}\nabla^{-2}\delta\) highlights the large-scale tidal or potential-like structure, \(\nabla_{ij}\delta\) gives the standard density-Hessian contrast, and the operators \(\nabla_{ij}\nabla^{2}\delta\) and \(\nabla_{ij}\nabla^{4}\delta\) progressively emphasize sharper and more local curvature features. Together, they provide a multiscale characterization of anisotropic web structure, in agreement with the spectral hierarchical web classification introduced by \citet[][]{2026arXiv260315834K}.}
\label{fig:cosmic-web2}
\end{figure}

Figure~\ref{fig:cosmic-web1} illustrates the basic sequence from an initially weakly structured field to an evolved nonlinear field, together with the associated potential representation. Figure~\ref{fig:cosmic-web2} shows the corresponding spectral hierarchy of Hessian-based web contrasts. These provide the geometric and multiscale backbone for the diagnostics that follow \citep[][]{Kitaura_2022,Coloma_2024,2026arXiv260315834K}.

\subsection{Density-dependent activation of anisotropy}
\label{subsec:anisotropic_growth}

A first quantitative illustration is provided by the conditional relation between density and anisotropy. The effective theory predicts that the anisotropy order parameter should remain small below a characteristic density threshold and become activated above it. In the minimal local picture this is expressed by the stationary branch
\begin{equation}
\left\langle Q^{(\ell)} \mid \dd^{(\ell)} \right\rangle
\simeq
\max\!\left[0,\,-\frac{a_{2,\ell}+c_\ell\dd^{(\ell)}}{2\beta_\ell}\right],
\label{eq:Qdelta_fit_main}
\end{equation}
or by an empirical generalization thereof.  {\color{black}Here we have used the linear expansion
\[
\alpha_\ell(\dd^{(\ell)})=a_{2,\ell}+c_\ell\dd^{(\ell)} ,
\]
so that the sign change of \(\alpha_\ell\) defines the onset of the
anisotropic branch.}

For this diagnostic we choose \(z=1\), which in cosmological terms corresponds to a stage where nonlinear clustering and web anisotropy are already clearly developed, but where the dynamics is still less dominated by the deepest multistreaming regime than at the present epoch, \(z=0\). This makes \(z=1\) a useful testing ground for the coarse-grained onset of anisotropic growth.

To connect the measured density--anisotropy relation with the effective free-energy picture, we fit the conditional mean anisotropy with the stationary Landau branch
\begin{equation}
Q_\star(\delta)=\max\!\left[0,\,s(\delta-\delta_c)\right],
\end{equation}
where \(\delta_c\) is the activation threshold and \(s>0\) is the growth slope of the anisotropic branch. Since the full measured \(\langle Q\mid\delta\rangle\) relation is not described equally well by this minimal form over the entire density range, we determine \(\delta_c\) and \(s\) from a restricted fit over moderate overdensities, where the thresholded linear branch provides a physically meaningful approximation. This yields \(\delta_c\simeq 0.2\) and \(s\simeq 13\), indicating that anisotropic clustering is activated already at moderate overdensity. 

This parametrization is equivalent to the minimal quadratic Landau form
\begin{equation}
f_{\rm loc}(\delta,Q)=f_{\rm iso}(\delta)+\alpha_Q(\delta)\,Q+\beta_Q Q^2,
\qquad
\alpha_Q(\delta)=a_{2Q}+c_Q\delta,
\end{equation}
with \(\beta_Q>0\), after identifying
\begin{equation}
Q_\star(\delta)=\max\!\left[0,\,-\frac{\alpha_Q(\delta)}{2\beta_Q}\right].
\end{equation}
For fixed \(\beta_Q\), the fitted branch parameters determine the effective coefficients through
\begin{equation}
a_{2Q}=2\beta_Q s\,\delta_c,
\qquad
c_Q=-2\beta_Q s.
\end{equation}
The corresponding minimized free-energy branch is
\begin{equation}
f_{\min}(\delta)=f_{\rm iso}(\delta)+\alpha_Q(\delta)\,Q_\star(\delta)+\beta_Q Q_\star(\delta)^2,
\end{equation}
and the free-energy gain relative to the isotropic solution is
\begin{equation}
\Delta f(\delta)\equiv f_{\min}(\delta)-f_{\rm iso}(\delta).
\end{equation}
By construction, \(\Delta f(\delta)=0\) where the isotropic branch remains preferred, while \(\Delta f(\delta)<0\) marks the density range in which anisotropy lowers the effective coarse-grained free-energy cost. In this sense, the fitted threshold \(\delta_c\) identifies the onset of anisotropic activation, whereas the later and more visible separation between \(f_{\min}\) and \(f_{\rm iso}\) reflects the increasing magnitude of the free-energy advantage at higher density.

\begin{figure}
\hspace{-0.5cm}
\begin{tabular}{cc}
\subfigure{\includegraphics[width=0.5\textwidth]{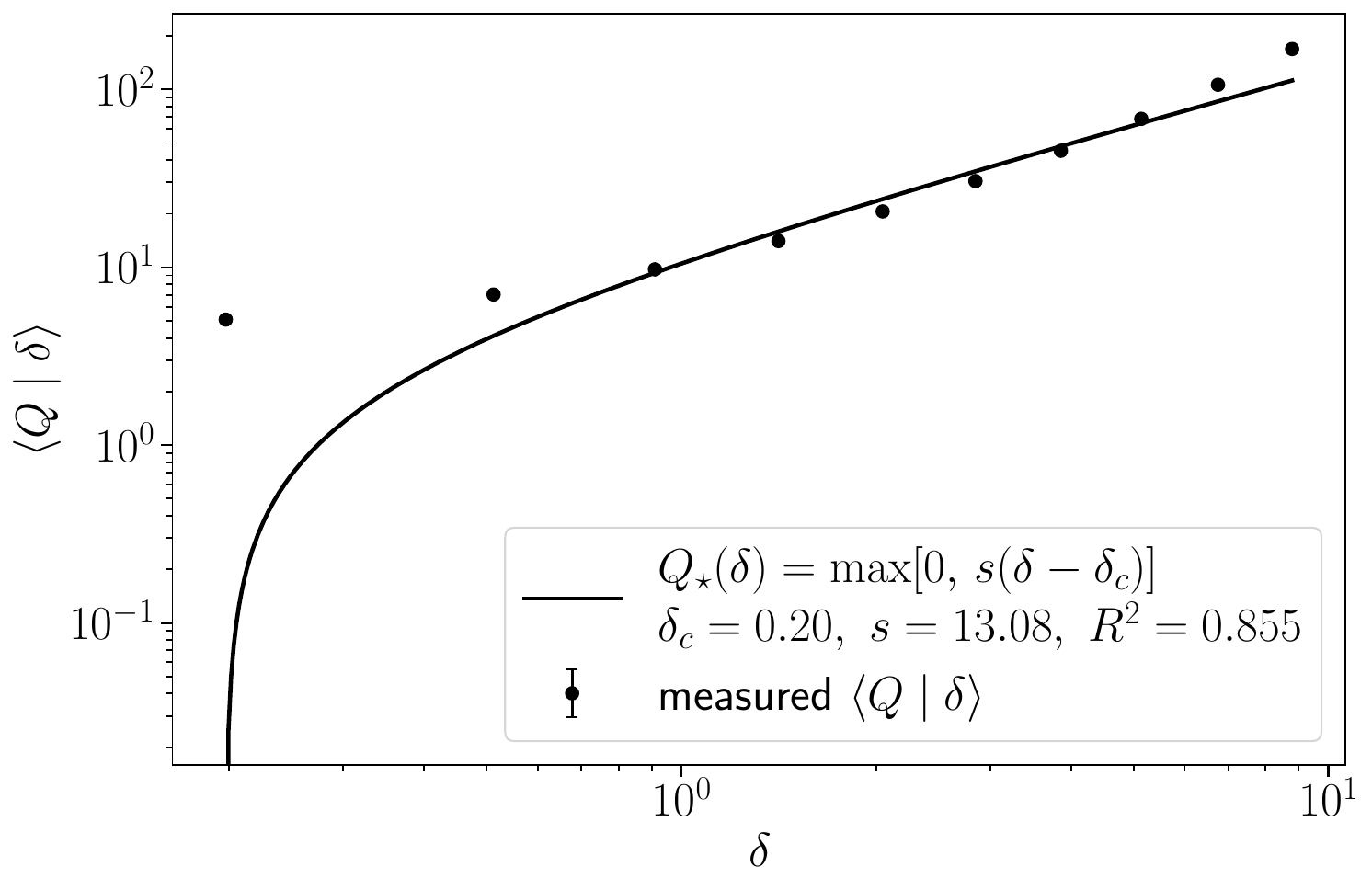}}
&
\subfigure{\includegraphics[width=0.5\textwidth]{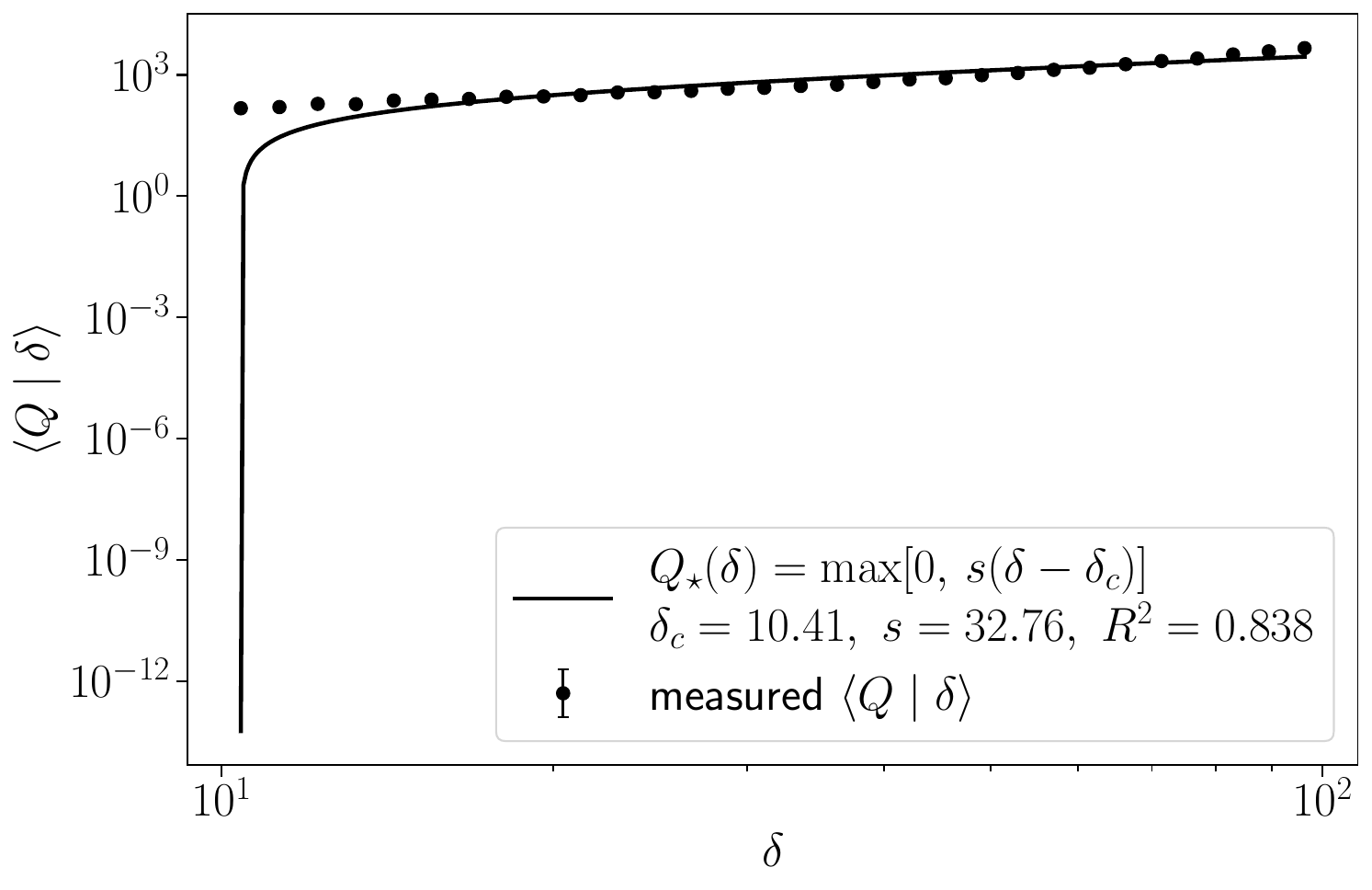}}
\vspace{-0.cm}
\\
\subfigure{\includegraphics[width=0.5\textwidth]{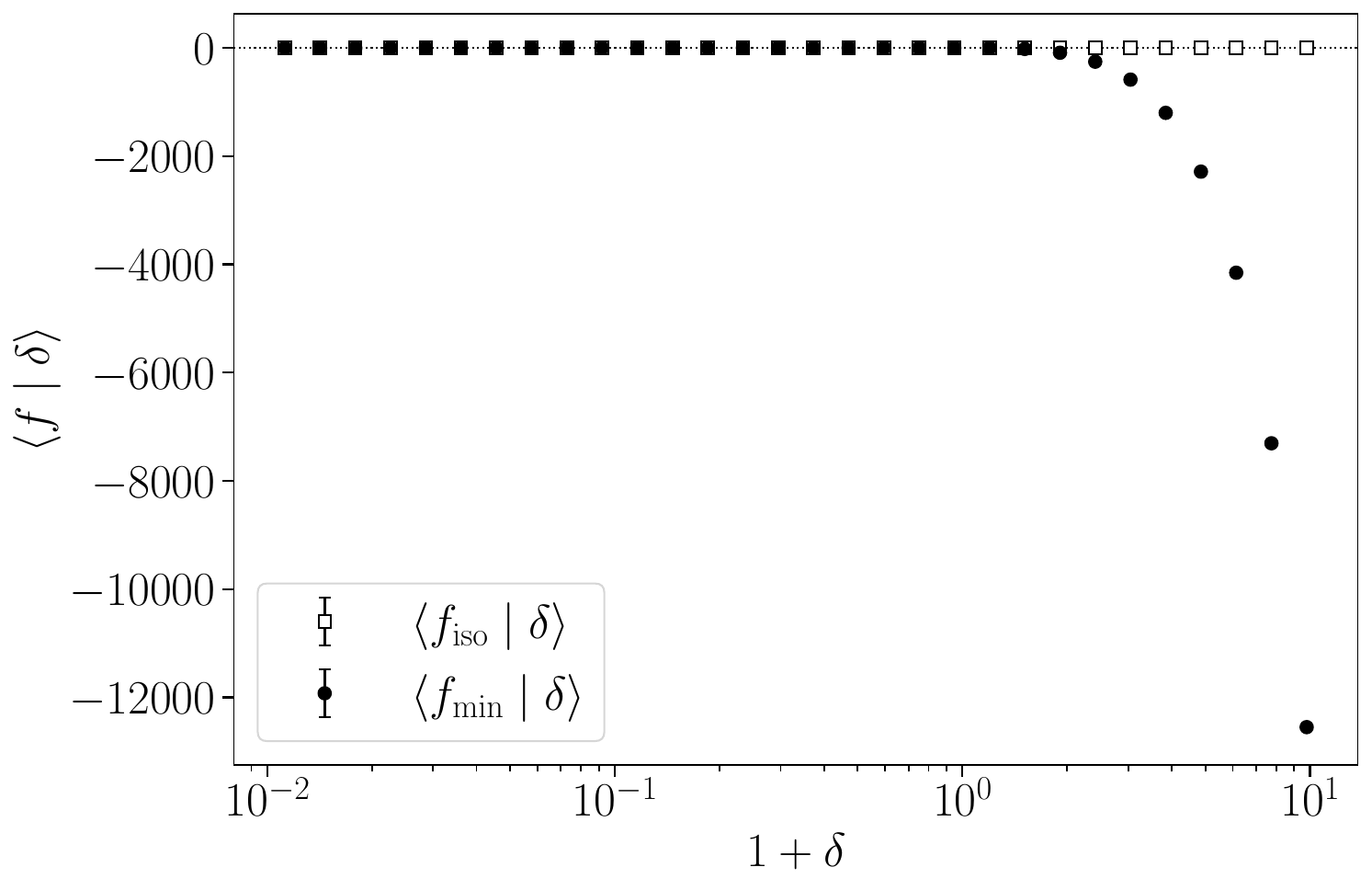}}
&
\subfigure{\includegraphics[width=0.5\textwidth]{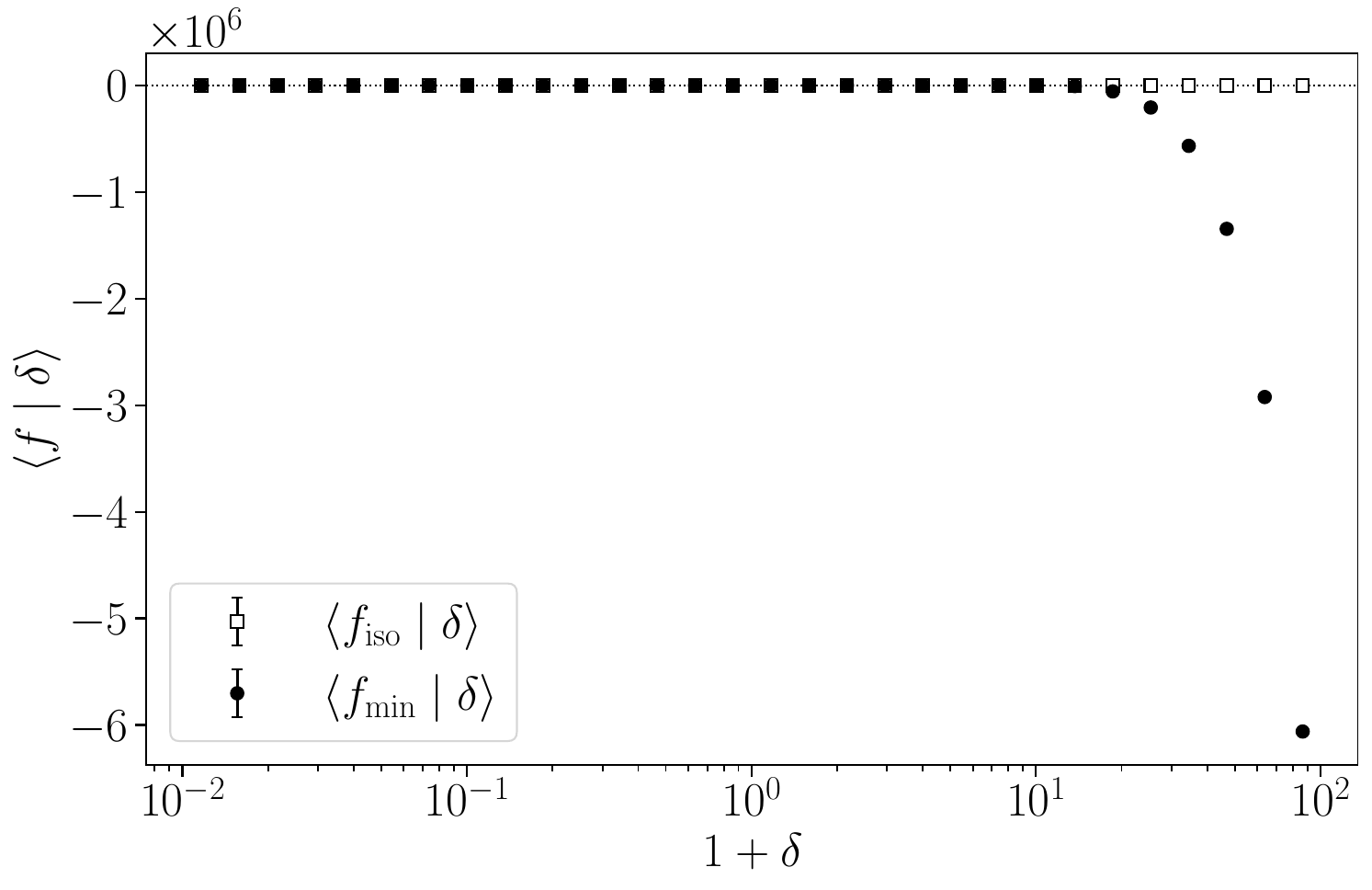}}
\end{tabular}
\vspace{-0.cm}
\caption{Comparison between the fitted anisotropy activation and the corresponding effective free-energy branches for two levels of the Hessian hierarchy. \textbf{Top left}: conditional mean anisotropy \(\langle Q\mid\delta\rangle\) for the \(k^{-2}\) level, corresponding to the tidal-field contraction built from \(\nabla_{ij}\nabla^{-2}\delta\), together with the restricted Landau-branch fit \(Q_\star(\delta)=\max[0,s(\delta-\delta_c)]\), yielding \(\delta_c\simeq 0.20\) and \(s\simeq 13.08\). \textbf{Top right}: the same for the \(k^{0}\) level, corresponding to the local curvature contraction built from \(\nabla_{ij}\delta\), with \(\delta_c\simeq 10.41\) and \(s\simeq 32.76\). \textbf{Bottom left}: comparison between the isotropic branch \(\langle f_{\rm iso}\mid\delta\rangle\) and the minimized anisotropic branch \(\langle f_{\min}\mid\delta\rangle\) derived from the \(k^{-2}\) fit. The two branches are nearly degenerate at low density, while the anisotropic branch becomes progressively favored beyond \(1+\delta\sim 2\)--\(3\). \textbf{Bottom right}: the corresponding branch comparison for the \(k^{0}\) level, showing that the free-energy separation becomes appreciable only later, around \(1+\delta\sim 10\). Together, the four panels illustrate a clear hierarchy: the nonlocal tidal field detects the onset of anisotropic organization already in the moderately nonlinear regime, whereas the more local density-curvature field responds strongly only in the denser, more nonlinear tail.}
\label{fig:landa-ginzburg-fits}
\end{figure}

The restricted fits shown in Figure \ref{fig:landa-ginzburg-fits} indicate that the onset of anisotropic activation depends on the operator used to define the coarse-grained anisotropy. For the \(k^{-2}\) level, associated with the tidal-field contraction \(\nabla_{ij}\nabla^{-2}\delta\), the fitted threshold is \(\delta_c\simeq 0.2\), implying that anisotropic clustering becomes relevant already at moderate overdensity and is clearly established by \(1+\delta\sim 2\)--\(3\). For the \(k^{0}\) level, associated with the curvature contraction \(\nabla_{ij}\delta\), the threshold shifts to \(\delta_c\simeq 10.4\), so that a strong anisotropic preference appears only later, around \(1+\delta\sim 10\). In this sense, the tidal field detects the onset of anisotropic organization earlier than the local curvature of the density field.
{\color{black}
This ordering is consistent with the physical roles of the two operators:
the \(k^{-2}\) tidal operator probes a broader nonlocal environment and
therefore registers coherent anisotropic forcing earlier, whereas the
local density-curvature operator responds most strongly only after collapse
has generated sharper, high-density structures. }

A more detailed web-environment decomposition of the conditional anisotropy curves is presented in Appendix~\ref{app:fitting}. These results show that the density--anisotropy relation is strongly environment dependent: the clearest and most extended activated branches are found in sheets and filaments, while knots and voids display less regular behavior, especially at higher hierarchical order.
 The numerical fits should be understood as a phenomenological realization of the general coarse-grained argument, not as its origin: the expectation of a thresholded anisotropic branch follows already from the Landau form itself.

\section{Discussion and conclusions}
\label{sec:discussion}

The main conceptual result of this work is that the apparent opposition between entropy growth and structure formation is largely an artifact of reduced descriptions. A coarse-grained spatial field may become more ordered as sheets, filaments, and knots emerge, and the corresponding spatial entropy may decrease. At the same time, the full phase-space state becomes more complex, because nonlinear evolution activates multistreaming, velocity dispersion, and higher-order structure that are invisible in a density-only description. The formation of complex morphology is therefore not in conflict with
entropy growth, but reflects the fact that projected spatial observables
capture only part of the richer dynamical information carried by
phase-space {\color{black} and by unresolved degrees of freedom.}

{\color{black}
The word "closed" should therefore be understood with care. In cosmology,
the Universe has no external environment with which it exchanges matter,
radiation or information, but it is not a fixed-volume Newtonian box with a
globally conserved total energy. In general relativity, energy-momentum
conservation is local and covariant, and a global conserved energy exists only in
special spacetime geometries. In an expanding
Friedmann--Lema\^itre--Robertson--Walker spacetime the energy in a
comoving volume is not generally constant. 
Radiation redshifts and vacuum-energy content grows with physical volume,
not because energy is injected from outside, but because the background
spacetime itself is dynamical. The entropy argument made here therefore
does not rely on global energy conservation, but on comparing different
coarse-grained descriptions of the same evolving system.

This qualification is even more important for non-cosmological examples.
A structural brain field, for instance, can only be treated as
approximately closed after choosing a mesoscopic variable such as an MRI
intensity, density proxy, or connectivity field. The brain is not a closed
thermodynamic system: it exchanges energy, matter and information with the
body and its environment. This is why the framework used here allows
source and sink terms in the generalized continuity equation, and why
Appendix~\ref{app:source_sink} discusses how the Jacobian relation is
modified when the transported quantity is not strictly conserved. Living
systems are an even clearer example of open self-organizing systems: they
maintain local order by consuming external free energy and exporting
entropy, as emphasized in non-equilibrium thermodynamics and in the theory
of dissipative structures associated with Prigogine and collaborators.
Such processes are conceptually related to self-organization, but modeling
metabolism, biological regulation, or life as a driven open system lies
outside the scope of the present work.

}

Within this perspective, shell crossing plays a special role. It is not only the geometric condition under which the Lagrangian--Eulerian map becomes singular in configuration space, but also the dynamical point at which the linear closure between density and velocity ceases to hold \citep[][]{2011JCAP...05..015S,Abel_2012,2016MNRAS.455.1115H,2016JCoPh.321..644S,2021A&A...647A..66C,2013ApJ...762..116Y}. Before shell crossing, the system admits a reduced effective description in terms of a single scalar mode or potential. After shell crossing, that compression is no longer possible: several streams may coexist at the same spatial location, the local state acquires nontrivial velocity structure, and a density-only characterization becomes incomplete. The same transition that generates the geometric skeleton of the cosmic web therefore also signals the growth of phase-space complexity.

A second important point is that anisotropy arises naturally in this framework. In the potential approximation, the deformation tensor reduces to a Hessian, and its eigenvalues determine the principal directions of local compression. Web-like morphology then follows from the generic splitting of these eigenvalues: collapse along one, two, or three directions produces sheets, filaments, and knots, respectively \citep[][]{Zeldovich_1970,Hahn_2007,Forero_2009,Bond_1996a,2012MNRAS.421L.137L,2026arXiv260315834K}. In this sense, anisotropic structure formation does not need to be imposed externally. It is already encoded in the geometry of the transport map and in the generic presence of local anisotropy in weakly structured initial fields.  In cosmology, this preference is ultimately seeded by primordial fluctuations themselves: a Gaussian random field is statistically almost never perfectly isotropic point by point, so gravitational amplification acts on an initially irregular Hessian landscape rather than on an exactly homogeneous background.

The statistical side of the framework complements this geometric picture. A Gaussian random field provides the maximum-entropy baseline consistent with prescribed two-point statistics, but nonlinear transport and nonlocal interaction drive the system away from this simple description. Mode coupling then generates higher-order correlations, such as skewness, kurtosis, bispectra, and trispectra, which constitute the statistical imprint of the same transition that geometrically produces complex morphology \citep[][]{Bernardeau_2002,Groth77,Baumgart91,Frieman94,Matarrese97,Verde98}. In this sense, web formation and non-Gaussianity should not be regarded as separate phenomena, but as two aspects of the same departure from the weakly structured Gaussian regime.

The coarse-grained Landau--Ginzburg viewpoint provides a useful way to organize these effects. Its role here is not to claim that all systems of interest literally thermalize to a common microscopic free energy, but to supply an effective mesoscopic language in which density, gradients, anisotropy invariants, and environmental dependence can be treated systematically \citep[][]{landau1965collected,Desjacques_2018,Kitaura_2022,Coloma_2024}. In this description, anisotropy becomes an order parameter, and the activation of anisotropic branches can be interpreted as the lowering of an effective coarse-grained free-energy cost. 

The numerical diagnostics presented in this work support this picture in the moderately nonlinear regime: the measured density--anisotropy relations exhibit clear activation thresholds, the activated branches depend on web environment, and the branch comparison indicates that the anisotropic solution acquires a lower effective coarse-grained free-energy cost once the corresponding anisotropy order parameter is activated.
In particular, the nonlocal tidal level becomes relevant already at moderate overdensity, whereas the more local density-curvature level becomes strongly favored only later.

From this viewpoint, the systems considered here may be interpreted within a broader class of self-organizing systems, namely systems in which coherent large-scale structure emerges spontaneously from internal dynamics acting on weakly organized initial conditions. In the present case, transport, anisotropic amplification, shell crossing, and nonlocal coupling collectively generate ordered morphology without the need for externally imposed patterning. What makes the cosmological case especially instructive is that this ordering can be tracked across levels of description: the coarse-grained spatial field becomes more structured, while the underlying phase-space state gains complexity through multistreaming and higher-order correlations. Self-organization is therefore not identified here with a decrease of total microscopic complexity, but rather with the spontaneous emergence of robust mesoscopic order under dynamical evolution.

This perspective also helps clarify a point that has appeared both in foundational discussions of gravitational entropy and in more speculative recent interpretations. Penrose famously emphasized the apparently paradoxical fact that the early Universe was extraordinarily smooth in its matter distribution and yet could not be regarded as a state of maximal physical entropy, precisely because gravity changes the meaning of what counts as equilibrium \citep[][]{1994JSP....77..217P}. Our results support this basic intuition in a coarse-grained dynamical setting. If one restricts attention to the projected one-point density field, the entropy associated with spatial occupancy can indeed decrease as clustering develops. But this decrease reflects only the growing inhomogeneity of a reduced description. Once the phase-space structure is included, shell crossing and multistreaming activate additional velocity-space complexity, so that the effective information required to characterize the macroscopic state increases. In this sense, the apparent ordering of the density field does not signal a fundamental entropy reduction or an intrinsic compression of information by the Universe \citep[][]{10.1063/5.0264945,2025PhRvX..15c1038C}; rather, it is the expected consequence of projecting an increasingly intricate phase-space dynamics onto a limited set of spatial observables.

Beyond the conceptual framework, the present work makes three concrete contributions. First, it formulates explicitly the distinction between projected spatial entropy and coarse-grained phase-space entropy. Second, it connects transport geometry, anisotropic collapse, and higher-order statistics within a single description linking Jacobian amplification, shell crossing, and non-Gaussianity. Third, it proposes a coarse-grained Landau--Ginzburg interpretation in which anisotropy is treated as an order parameter and web-like morphology as the activation of energetically favored anisotropic branches.

Several limitations and extensions remain. The present effective description is intentionally minimal, and the free-energy truncation used here is not expected to capture the deepest multistreaming regime in a quantitatively complete way. Likewise, vorticity, non-potential transport, explicit memory effects, and more general nonlocal kernels may become important in strongly nonlinear systems \citep[][]{Catelan_1995,2011JCAP...04..032M,2012JCAP...01..019P,Rampf_2012,2017PhRvD..95f3527C,2005A&A...438..443B,2026arXiv260313106K}. On the numerical side, higher-resolution phase-space methods could help clarify more sharply the relation between coarse-grained anisotropy diagnostics and the underlying fine-grained multistream structure \citep[][]{2013ApJ...762..116Y,2016JCoPh.321..644S,2020MNRAS.493.2765S,2021A&A...647A..66C,2024MNRAS.52710802O}. More broadly, the extent to which analogous operator hierarchies and free-energy-like descriptions apply outside cosmology, for example in connectome-like or other spatially extended complex systems, remains an open and promising direction \citep[][]{ROSELL2025121500,2024PhLRv..48...47P,2025NatSR..1542309K}.

{\color{black} In this sense, complex web formation can be viewed as a general spontaneous
breaking mechanism of local isotropy, through which nonlinear transport
selects preferred directions and favours anisotropic coarse-grained
structures.}
The main message is therefore a unifying one. Transport, shell crossing, multistreaming, anisotropy, higher-order correlations, and effective free-energy minimization are not disconnected ingredients. They are different manifestations of the same transition by which an initially weakly structured field leaves the regime of Gaussian, density-only, effectively single-stream organization and enters a genuinely multiscale, anisotropic, and nonlocal phase-space state. Although cosmological structure formation provides the main realization considered here, the broader aim of the paper is conceptual. The framework is intended as a common mesoscopic language for systems in which weakly structured initial states evolve into complex anisotropic morphology through transport, nonlocality, and nonlinear amplification.

\newpage

\appendix

\section{From phase-space dynamics to the continuity equation}
\label{app:continuity}

In the collisionless case, the microscopic state of the system is described by the phase-space distribution function \(f(t,\vx,\vv)\) \citep[][]{Bernardeau_2002,2010gfe..book.....M,Angulo_2022}. Its evolution is governed by the Vlasov equation,
\begin{equation}
\partial_t f
+
\vv\cdot\nabla_{\vx} f
+
\vg\cdot\nabla_{\vv} f
=
0,
\label{eq:vlasov_app}
\end{equation}
where \(\vg(\vx,t)\) is the acceleration field. The mass density and mean velocity are the lowest moments of \(f\),
\begin{equation}
\rr(\vx,t)=m\int \diff^n v\, f(t,\vx,\vv),
\qquad
\rr(\vx,t)\,\vu(\vx,t)=m\int \diff^n v\, f(t,\vx,\vv)\,\vv.
\label{eq:density_velocity_app}
\end{equation}

Integrating Eq.~\eqref{eq:vlasov_app} over velocity space gives
\begin{equation}
{\color{black} m}\int \partial_t f\,\diff^n v
+
{\color{black} m}\int \vv\cdot\nabla_\vx f\,\diff^n v
+
{\color{black} m}\int \vg\cdot\nabla_\vv f\,\diff^n v
=
0.
\end{equation}
The first term yields \(\partial_t \rr\), and the second yields \(\nabla_\vx\cdot(\rr \vu)\). For the third term, assuming sufficiently rapid decay of \(f\) as \(|\vv|\to\infty\), the boundary contribution in velocity space vanishes. One therefore obtains
\begin{equation}
\partial_t \rr+\nabla_\vx\cdot(\rr \vu)=0.
\label{eq:continuity_appendix_final}
\end{equation}
This is the continuity equation quoted in the main text.

If local production or loss is present, the same balance-law logic leads to
\begin{equation}
\partial_t \rr+\nabla_\vx\cdot(\rr \vu)=\sigma,
\label{eq:continuity_source_appendix_final}
\end{equation}
with \(\sigma(\vx,t)\) a source or sink term.

\section{Projected spatial entropy and its decrease under clustering}
\label{app:spatial_entropy}

To make explicit the entropy decrease of a coarse-grained spatial field, consider a partition of the domain into \(N_c\) cells of equal volume, and let \(\rr_i\) be the coarse-grained density in cell \(i\). Define the normalized spatial probabilities
\begin{equation}
p_i=\frac{\rr_i}{\sum_{j=1}^{N_c}\rr_j},
\qquad
\sum_{i=1}^{N_c}p_i=1.
\end{equation}
The corresponding spatial Shannon entropy is
\begin{equation}
S_x=-\sum_{i=1}^{N_c} p_i\ln p_i.
\label{eq:spatial_entropy_app}
\end{equation}
For a perfectly homogeneous state, \(p_i=1/N_c\), and therefore
\begin{equation}
S_x^{\max}=\ln N_c.
\end{equation}

Now consider small deviations from homogeneity,
\begin{equation}
p_i=\frac{1}{N_c}+\epsilon_i,
\qquad
\sum_i \epsilon_i=0.
\end{equation}
Expanding Eq.~\eqref{eq:spatial_entropy_app} to second order gives
\begin{equation}
S_x
=
\ln N_c
-
\frac{N_c}{2}\sum_{i=1}^{N_c}\epsilon_i^2
+
\mathcal O(\epsilon^3).
\label{eq:spatial_entropy_expand_app}
\end{equation}
Thus any departure from homogeneity lowers the coarse-grained spatial entropy \citep[][]{cover2012elements,shannon}.

If one writes the density as \(\rr(\vx)=\rrb\,[1+\dd(\vx)]\) with \(|\dd|\ll 1\), then
\begin{equation}
S_x
\approx
\ln N_c
-
\frac{1}{2N_c}\sum_i \dd_i^2,
\label{eq:spatial_entropy_delta_app}
\end{equation}
showing explicitly that the spatial entropy decreases as the variance of the coarse-grained density field increases.

This is the precise sense in which an initially homogeneous noisy field can evolve toward a more ordered spatial morphology while the associated projected spatial entropy decreases.

\section{Phase-space entropy decomposition and multistreaming}
\label{app:phase_space_entropy}

The decrease of the projected spatial entropy does not capture the full information content of the evolving system. To see this, partition phase-space into \(N_x\) spatial cells and \(N_v\) velocity cells, and let \(m_{i\alpha}\) be the mass in spatial cell \(i\) and velocity cell \(\alpha\). Define the normalized phase-space probabilities
\begin{equation}
\pi_{i\alpha}
=
\frac{m_{i\alpha}}{\sum_{j,\beta}m_{j\beta}},
\qquad
\sum_{i,\alpha}\pi_{i\alpha}=1.
\end{equation}
The corresponding coarse-grained phase-space entropy is
\begin{equation}
S_{\rm ps}
=
-\sum_{i,\alpha}\pi_{i\alpha}\ln\pi_{i\alpha}.
\label{eq:phase_space_entropy_app}
\end{equation}

The spatial probabilities are the marginals
\begin{equation}
p_i=\sum_\alpha \pi_{i\alpha},
\qquad
S_x=-\sum_i p_i\ln p_i.
\label{eq:spatial_marginal_app}
\end{equation}
The conditional velocity probabilities are
\begin{equation}
w_{\alpha|i}=\frac{\pi_{i\alpha}}{p_i},
\qquad
\sum_\alpha w_{\alpha|i}=1.
\end{equation}
Using \(\pi_{i\alpha}=p_i\,w_{\alpha|i}\), one finds the exact decomposition
\begin{equation}
S_{\rm ps}=S_x+S_{v|x},
\label{eq:entropy_chain_rule_app}
\end{equation}
where
\begin{equation}
S_{v|x}
=
-\sum_i p_i\sum_\alpha w_{\alpha|i}\ln w_{\alpha|i}
\label{eq:conditional_velocity_entropy_app}
\end{equation}
is the conditional entropy of velocities given position.

In the single-stream regime, each spatial cell carries effectively a unique coarse-grained velocity state,
\begin{equation}
w_{\alpha|i}=\Kronecker_{\alpha,\alpha_i},
\end{equation}
and therefore
\begin{equation}
S_{v|x}=0,
\qquad
S_{\rm ps}=S_x.
\end{equation}
After shell crossing, several streams with distinct velocities may coexist at the same spatial location, so \(w_{\alpha|i}\) has support on multiple velocity bins and
\begin{equation}
S_{v|x}>0.
\end{equation}
Hence
\begin{equation}
S_{\rm ps}>S_x.
\end{equation}

This decomposition makes explicit the central distinction used throughout the paper: the projected spatial entropy may decrease as clustering develops, while the full coarse-grained phase-space entropy grows because multistreaming activates additional velocity-space complexity \citep[][]{Abel_2012,2011JCAP...05..015S,2014MNRAS.437.3442H}.

\section{Jacobian amplification with source or sink terms}
\label{app:source_sink}

Let the transport map be written as \(\vx=\vx(\vq,t)\), with Jacobian
\begin{equation}
J(\vq,t)=\det\!\left(\frac{\partial \vx}{\partial \vq}\right).
\end{equation}
For conserved transport, mass conservation gives
\begin{equation}
\rr(\vx,t)\,\diff^n x=\rr(\vq,t_0)\,\diff^n q,
\end{equation}
hence
\begin{equation}
\rr(\vx(\vq,t),t)=\rr_0(\vq)\,J(\vq,t)^{-1}.
\label{eq:rho_J_conservative_app}
\end{equation}

More generally, let the continuity equation include a source or sink term,
\begin{equation}
\partial_t \rr+\nabla_\vx\cdot(\rr\vu)=\sigma.
\label{eq:continuity_source_app}
\end{equation}
Using Jacobi's formula, one finds along a trajectory
\begin{equation}
\frac{\diff J}{\diff t}=J\,\nabla_\vx\cdot\vu.
\end{equation}
Combining this with Eq.~\eqref{eq:continuity_source_app} gives
\begin{equation}
\frac{\diff}{\diff t}(\rr J)=J\,\sigma.
\label{eq:rhoJ_source_app}
\end{equation}

A particularly simple case is a sink proportional to the density,
\begin{equation}
\sigma(\vx,t)=-\gamma(\vx,t)\,\rr(\vx,t),
\qquad \gamma\ge 0.
\end{equation}
Then Eq.~\eqref{eq:rhoJ_source_app} integrates to
\begin{equation}
\rr(\vx(\vq,t),t)
=
\rr_0(\vq)\,J(\vq,t)^{-1}e^{-\ee(\vq,t)},
\label{eq:rho_source_sink_final_app}
\end{equation}
with
\begin{equation}
\ee(\vq,t)=\int_{t_0}^{t}\gamma(\vx(\vq,s),s)\,\diff s.
\end{equation}
Thus source and sink terms modify the purely Jacobian amplification by an additional cumulative exponential factor.

\section{Maximum-entropy Gaussian fields and higher-order generalization}
\label{app:maxent}

The Shannon entropy of a probability density \(p(x)\) is
\begin{equation}
S[p]=-\int \diff x\, p(x)\ln p(x).
\end{equation}
Maximizing \(S[p]\) subject to normalization and fixed first and second moments,
\begin{equation}
\int \diff x\,p(x)=1,
\qquad
\int \diff x\,x\,p(x)=\mu,
\qquad
\int \diff x\,x^2p(x)=m_2,
\end{equation}
gives, by the method of Lagrange multipliers,
\begin{equation}
p(x)\propto \exp\!\left(-\lambda_1 x-\lambda_2 x^2\right).
\end{equation}
Completing the square yields the Gaussian form
\begin{equation}
p(x)=\frac{1}{\sqrt{2\pi\sigma^2}}
\exp\!\left[-\frac{(x-\mu)^2}{2\sigma^2}\right].
\label{eq:gaussian_maxent_app}
\end{equation}

For a field \(\varphi(\vx)\), the corresponding maximum-entropy ensemble with fixed mean and covariance is the Gaussian functional
\begin{equation}
P[\varphi]\propto
\exp\left[
-\frac12
\int \diff^n x\,\diff^n y\;
(\varphi(\vx)-\mu(\vx))\,C^{-1}(\vx,\vy)\,(\varphi(\vy)-\mu(\vy))
\right].
\label{eq:gaussian_functional_app}
\end{equation}
This is the statistical baseline used in the main text \citep[][]{1957PhRv..106..620J,cover2012elements,adler1981geometry}.

Once higher-order moments are also constrained, the exponent is generalized from a quadratic to a higher-order polynomial. If one fixes normalization and the first few moments up to fourth order, the maximum-entropy solution takes the schematic form
\begin{equation}
p(x)\propto
\exp\!\left(
-\lambda_1 x
-\lambda_2 x^2
-\lambda_3 x^3
-\lambda_4 x^4
-\cdots
\right).
\label{eq:nongaussian_maxent_app}
\end{equation}
Thus nonzero higher cumulants correspond formally to higher-order corrections to the Gaussian exponent. In particular, a nonzero third moment or bispectrum induces a cubic correction, while a nonzero fourth moment or trispectrum induces a quartic correction. In practice, stabilizing even-order terms are required for normalizability on unbounded support \citep[][]{cover2012elements,1957PhRv..106..620J}.

\section{Second-order mode coupling and non-Gaussianity}
\label{app:ept}

The key statistical consequence of nonlinear transport is that Fourier modes cease to evolve independently. In standard perturbative language, the field may be expanded as
\begin{equation}
\dd(\vk)=\dd^{(1)}(\vk)+\dd^{(2)}(\vk)+\cdots.
\end{equation}
At second order, one obtains the generic convolution structure
\begin{equation}
\dd^{(2)}(\vk)
=
\int \frac{\diff^n p}{(2\pi)^n}\,
F_2(\vp,\vk-\vp)\,
\dd^{(1)}(\vp)\dd^{(1)}(\vk-\vp),
\label{eq:F2_generic_app}
\end{equation}
where \(F_2\) is the second-order mode-coupling kernel \citep[][]{Bernardeau_2002} {\color{black} and where we have set $\varphi=\dd^{(1)}$ in Eq.~\ref{eq:mode_coupling_main}}.

In the standard pressureless cosmological case, the Einstein--de Sitter kernel is
\begin{equation}
F_2(\vk_1,\vk_2)
=
\frac{5}{7}
+
\frac{1}{2}
\frac{\vk_1\cdot\vk_2}{k_1 k_2}
\left(
\frac{k_1}{k_2}
+
\frac{k_2}{k_1}
\right)
+
\frac{2}{7}
\left(
\frac{\vk_1\cdot\vk_2}{k_1 k_2}
\right)^2.
\label{eq:F2_EdS_app}
\end{equation}
The exact coefficients are model dependent, but the convolution structure itself is generic: the second-order field at \(\vk\) is sourced by pairs of lower-order modes, and therefore the evolved field is no longer characterized by the power spectrum alone.

This is the origin of connected higher-order correlators. In Fourier space,
\begin{equation}
\langle \dd(\vk_1)\dd(\vk_2)\rangle
=
(2\pi)^n\Dirac(\vk_1+\vk_2)\,P(\vk_1),
\end{equation}
\begin{equation}
\langle \dd(\vk_1)\dd(\vk_2)\dd(\vk_3)\rangle_{\rm c}
=
(2\pi)^n\Dirac(\vk_1+\vk_2+\vk_3)\,B(\vk_1,\vk_2,\vk_3),
\end{equation}
and similarly for the trispectrum and higher connected correlators. Thus mode coupling provides the direct statistical bridge from Gaussian baselines to non-Gaussian structure \citep[][]{Bernardeau_2002,Groth77,Baumgart91,Frieman94,Matarrese97,Verde98}.

\section{Why coarse-graining leads to a free-energy functional}
\label{app:LG_motivation}

The Landau--Ginzburg viewpoint used in the main text is motivated by coarse-graining. At the semi-microscopic level, one may define a constrained partition function for a coarse-grained field \(\phi(\vx)\),
\begin{equation}
Z[\phi]
=
\int \mathcal D\sigma\,
\Dirac\!\bigl(\phi-\phi_\ell[\sigma]\bigr)\,
e^{-\beta H[\sigma]},
\label{eq:Z_phi_app}
\end{equation}
where \(\sigma\) denotes the microscopic degrees of freedom and \(\phi_\ell[\sigma]\) the corresponding coarse-grained field at resolution \(\ell\). The associated coarse-grained free energy is
\begin{equation}
F_{\rm cg}[\phi]=-k_B T\ln Z[\phi].
\label{eq:Fcg_app}
\end{equation}

The key point is that integrating over unresolved microscopic degrees of freedom naturally combines energetic and entropic contributions into a mesoscopic functional \citep[][]{landau1965collected}. If the coarse-grained field varies slowly in space, that functional can be expanded in the most general local scalar operators compatible with the symmetries of the problem. In its simplest form this gives a Landau--Ginzburg-type expression,
\begin{equation}
F_{\rm LG}[\phi]
=
\int \diff^n x\,
\left[
a+b\,\phi^2+c\,\phi^4+d\,|\nabla\phi|^2+\cdots
\right].
\label{eq:LG_generic_app}
\end{equation}

In the present work the role of the coarse-grained variables is played not by a single scalar field alone, but by the density contrast \(\dd\) together with a hierarchy of Hessian-like tensors \(E^{(\ell)}_{ij}\). The effective functional is therefore written in terms of density amplitudes, gradients, and anisotropy-sensitive invariants. Its purpose is not to claim a unique microscopic Hamiltonian for all systems of interest, but to provide a systematic operator expansion for morphology after coarse-graining \citep[][]{landau1965collected}.

\section{Anisotropy order parameter and stationary branch}
\label{app:anisotropy_stationarity}

For each hierarchy level \(\ell\) \citep[][]{2026arXiv260315834K}, define the quadratic anisotropy invariant
\begin{equation}
{\color{black}  Q^{(\ell)}
=
\left(E^{(\ell)}_{ij}
-\frac{1}{3}\delta^{\rm K}_{ij}E^{(\ell)}_{kk}\right)
\left(E^{(\ell)}_{ij}
-\frac{1}{3}\delta^{\rm K}_{ij}E^{(\ell)}_{kk}\right)}\ge 0.
\label{eq:Q_app}
\end{equation}
The different operators associated with the hierarchy levels \(\ell\) are indicated in Figure~\ref{fig:cosmic-web2}.
By construction, \(Q^{(\ell)}=0\) only for locally isotropic configurations and becomes positive as the eigenvalues of \(E^{(\ell)}_{ij}\) split.

A minimal local effective density may then be written as
\begin{equation}
f_{\rm loc}(\dd,\{Q^{(\ell)}\})
=
f_\dd(\dd)
+
\sum_\ell
\left[
\alpha_\ell(\dd)\,Q^{(\ell)}
+
\beta_\ell \bigl(Q^{(\ell)}\bigr)^2
\right],
\qquad \beta_\ell>0.
\label{eq:floc_stationary_app}
\end{equation}
Minimizing \(f_{\rm loc}\) with respect to \(Q^{(\ell)}\) gives
\begin{equation}
\frac{\partial f_{\rm loc}}{\partial Q^{(\ell)}}=
\alpha_\ell(\dd)+2\beta_\ell Q^{(\ell)}=0,
\end{equation}
so the formal stationary branch is
\begin{equation}
Q^{(\ell)}_\star(\dd)=
-\frac{\alpha_\ell(\dd)}{2\beta_\ell}.
\end{equation}
Since \(Q^{(\ell)}\ge 0\), the physical branch is
\begin{equation}
Q^{(\ell)}_\star(\dd)
=
\max\!\left[
0,\,
-\frac{\alpha_\ell(\dd)}{2\beta_\ell}
\right].
\label{eq:Qstar_stationary_app}
\end{equation}

The usual Landau interpretation follows immediately. If \(\alpha_\ell(\dd)>0\), the isotropic branch \(Q^{(\ell)}=0\) is locally preferred. If \(\alpha_\ell(\dd)<0\), the isotropic branch ceases to minimize the local free-energy density and an anisotropic branch with \(Q^{(\ell)}_\star>0\) is activated. This is the basis of the density-dependent anisotropic activation discussed in the main text \citep[][]{landau1965collected}.

\subsection{Eigenvalue interpretation}

If \(\mu_1^{(\ell)},\mu_2^{(\ell)},\mu_3^{(\ell)}\) are the eigenvalues of \(E^{(\ell)}_{ij}\), and
\begin{equation}
\bar\mu^{(\ell)}=\frac{\mu_1^{(\ell)}+\mu_2^{(\ell)}+\mu_3^{(\ell)}}{3},
\end{equation}
then
\begin{equation}
{\color{black} {Q}}^{(\ell)}
=
\sum_{i=1}^3
\left(\mu_i^{(\ell)}-\bar\mu^{(\ell)}\right)^2\,,
\label{eq:Q_eigen_app}
\end{equation}
{\color{black} where ${Q}^{(\ell)}=\tilde Q^{(\ell)}-3\bar\mu^2$}.
Thus \(Q^{(\ell)}\) is precisely a scalar measure of eigenvalue splitting, and therefore of anisotropy \citep[][]{McDonald_2009,Chan_2012,Desjacques_2018,2018MNRAS.476.3631P,Kitaura_2022,Coloma_2024,Sinigaglia_2024}.

\section{Why anisotropy  is expected to become favored}
\label{app:anisotropic_branch_growth}

The appearance of an anisotropic branch in the effective coarse-grained description is not introduced here as a purely phenomenological fitting ansatz. It is already suggested by the structure of the growth equations and by the geometry of transport in the growing-mode regime.

Rather than committing to a specific closure from the outset, it is useful to adopt the standard effective-theory viewpoint in which the large-scale dynamics is written as the continuity--Euler--Poisson system supplemented by an effective stress or acceleration term that admits a symmetry-based derivative expansion \citep[][]{Bernardeau_2002,2010gfe..book.....M}. In this formulation, the additional contribution is organized as an isotropic series in increasingly higher spatial derivatives of the density field. The particular hierarchy of Hessian-like operators used in the present work may then be viewed as a concrete realization of this general expansion, reordered according to the spectral operator levels introduced by {\color{black} Kitaura \& Sinigaglia 2026} \cite[][]{2026arXiv260315834K}.

In the growing-mode regime, the Zel'dovich approximation provides the simplest separable realization of this idea,
\begin{equation}
\vPsi(\vq,t)=D(t)\,\nabla\phi(\vq),
\end{equation}
so that the deformation tensor becomes
\begin{equation}
{\cal D}_{ij}(\vq,t)=D(t)\,\phi_{,ij}(\vq).
\end{equation}
If \(\mu_1\leq \mu_2\leq \mu_3\) are the eigenvalues of \(\phi_{,ij}\), then
\begin{equation}
J(\vq,t)=\prod_{a=1}^3 \bigl(1+D(t)\mu_a(\vq)\bigr),
\qquad
1+\dd(\vq,t)=\prod_{a=1}^3 \bigl(1+D(t)\mu_a(\vq)\bigr)^{-1}.
\end{equation}
The key point is that the growing mode amplifies the initial deformation tensor multiplicatively, so any primordial eigenvalue splitting is carried forward and sharpened rather than averaged away. Collapse occurs first along the most compressive direction, namely the eigenvalue that reaches
\begin{equation}
1+D(t)\mu_a(\vq)=0
\end{equation}
first. This yields the familiar sequence sheet \(\rightarrow\) filament \(\rightarrow\) knot. Exact isotropic collapse would require degenerate eigenvalues and is therefore non-generic. In a Gaussian initial field, such degeneracy has vanishing probability; already the classic analysis of {\color{black} Doroshkevich 1970}  \citep[][]{1970Afz.....6..581D} showed that the initial deformation field of a Gaussian random realization is generically triaxial.

From the coarse-grained point of view, this means that the growth dynamics naturally drives the system away from the isotropic branch \(Q=0\) and toward configurations with nonzero eigenvalue splitting, \(Q>0\). The anisotropic branch should therefore be interpreted as the effective continuation of the physically realized growing solution once the system leaves the weakly structured regime. In this sense, the sign change of the coefficient multiplying \(Q\) in the Landau--Ginzburg functional is not arbitrary: it is the coarse-grained encoding of the fact that attractive growth generically amplifies pre-existing anisotropic deformations, while perfectly isotropic collapse corresponds to a special degenerate case.

The quantitative measurements presented in the main text should therefore be understood only as an illustration of this more general logic. They verify, in a concrete cosmological realization, that the nonlocal tidal level becomes relevant earlier and that more local curvature-sensitive levels become important later, in agreement with the general hierarchy of growth and with the geometric sequence of anisotropic collapse.

\section{Fitting procedure for the anisotropy curves}
\label{app:fitting}

\begin{figure}
\vspace{0cm}
\centering
\includegraphics[width=0.6\textwidth]{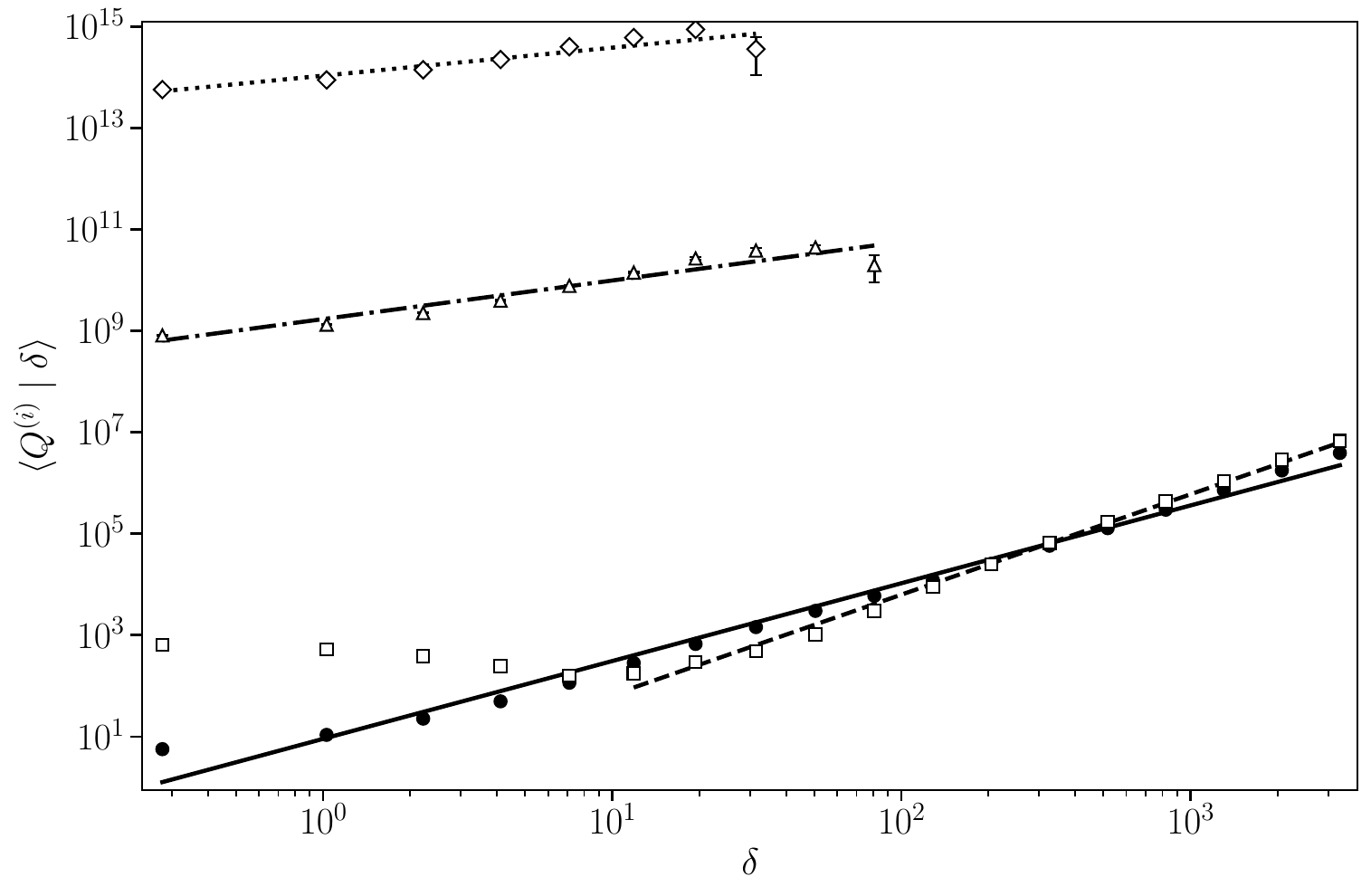}
\put(-170,100){\footnotesize knots $\cup$ filaments $\cup$ sheets $\cup$ voids}
\caption{Conditional mean tidal anisotropy \(\langle Q^{(i)} \mid \dd \rangle\) for the four hierarchical levels \(i=1,\dots,4\), measured from the generated fields. Binning is performed uniformly in \(\log(1+\dd)\). Symbols denote the measured averages and lines the best-fitting shifted power laws \(C_i(\dd-\dd_{c,i})^{p_i}\). In all cases, anisotropy is negligible below a characteristic density threshold and rises rapidly above it, consistent with the activation of anisotropic branches in the coarse-grained description.}
\label{fig:landau-ginzburg}
\end{figure}

\begin{figure}
\vspace{0cm}
\centering
\begin{tabular}{cc}
\hspace{-0.75cm}
\subfigure{\includegraphics[width=0.5\textwidth]{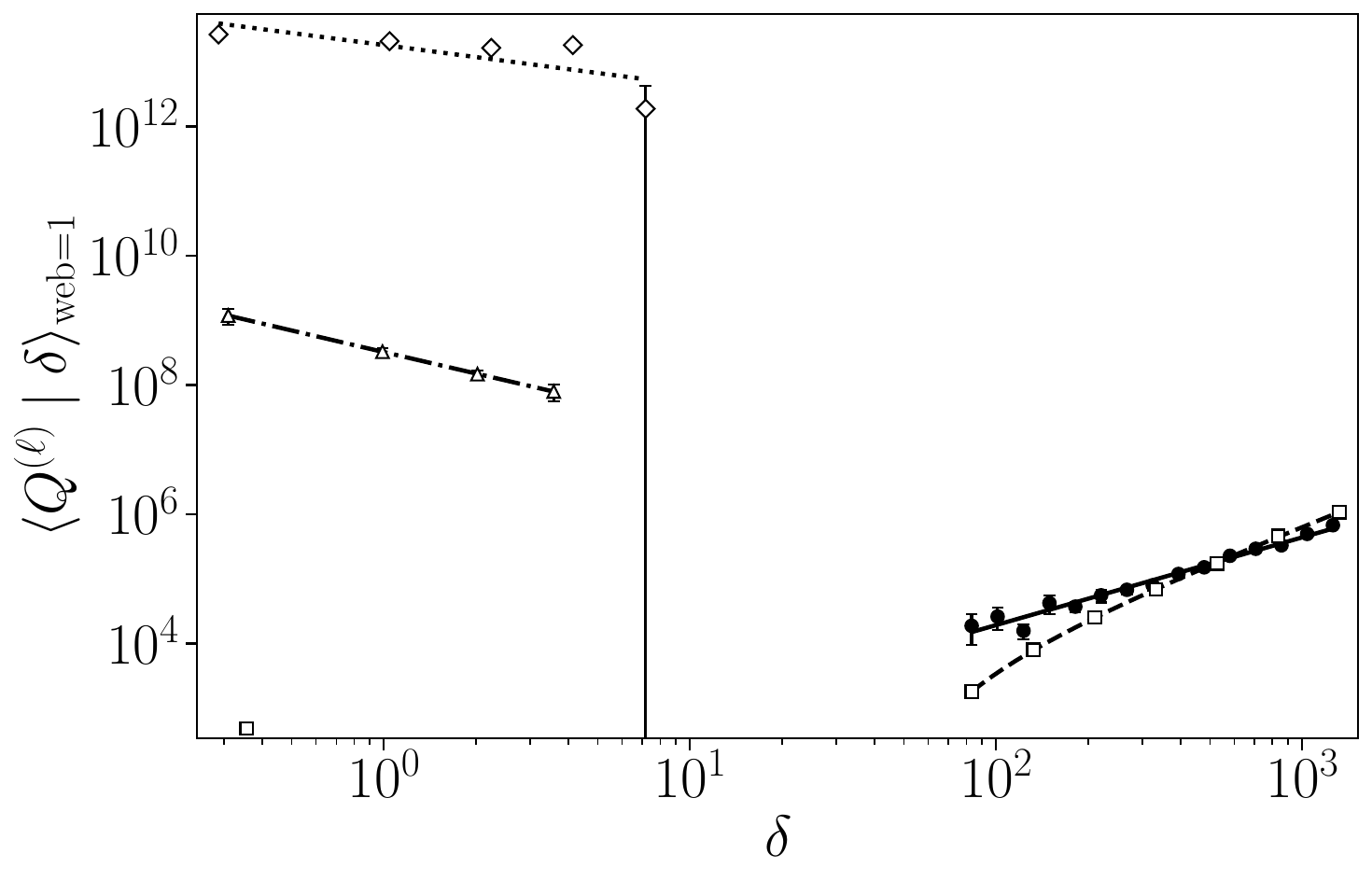}}
\put(-90,80){\footnotesize knots}
\hspace{-0.5cm}
&
\subfigure{\includegraphics[width=0.5\textwidth]{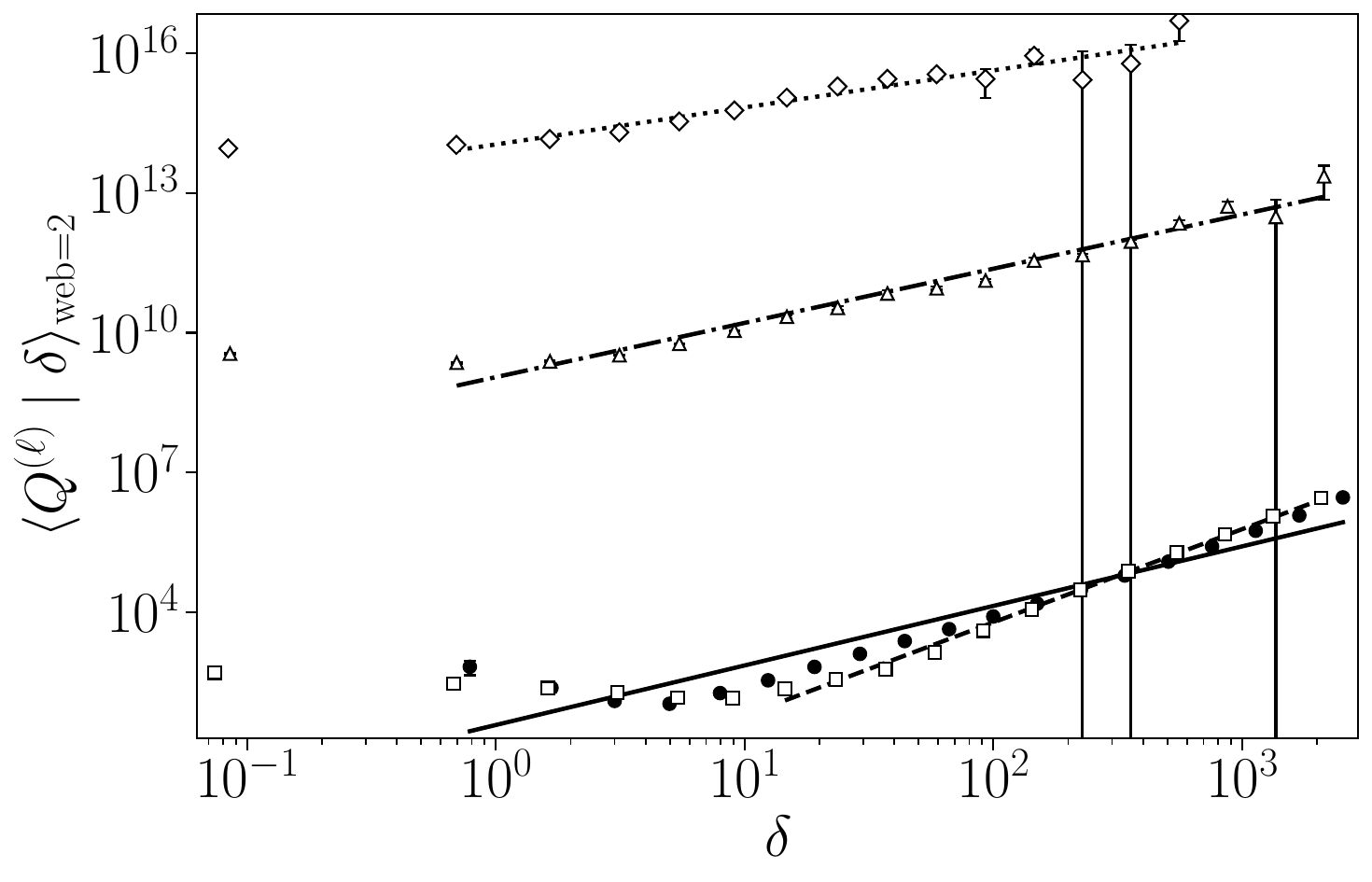}}
\put(-90,80){\footnotesize filaments}
\vspace{-.0cm}
\\
\hspace{-0.55cm}
\subfigure{\includegraphics[width=0.5\textwidth]{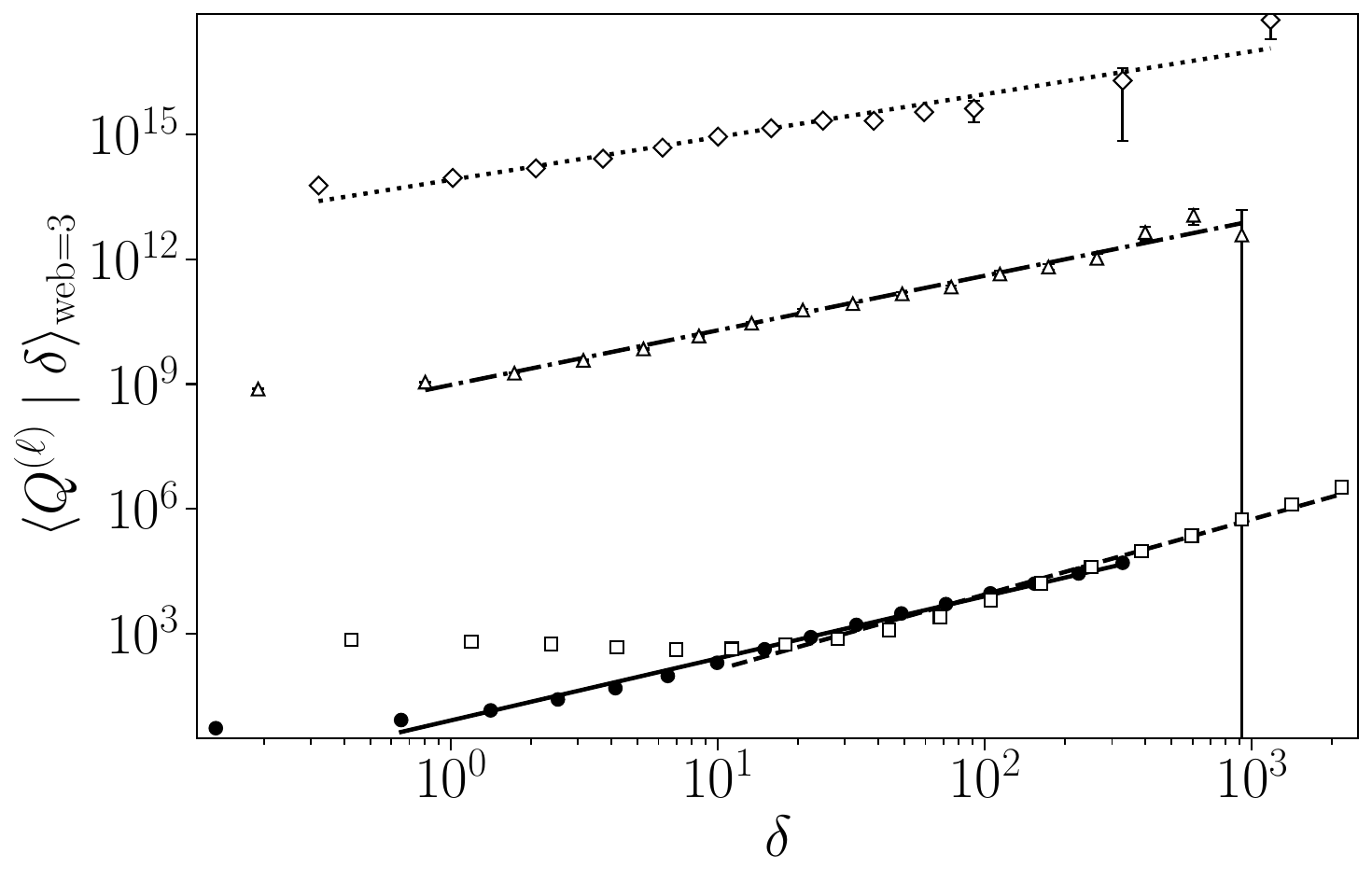}}
\put(-90,80){\footnotesize sheets}
\hspace{-0.3cm}
&
\hspace{-0.1cm}
\subfigure{\includegraphics[width=0.5\textwidth]{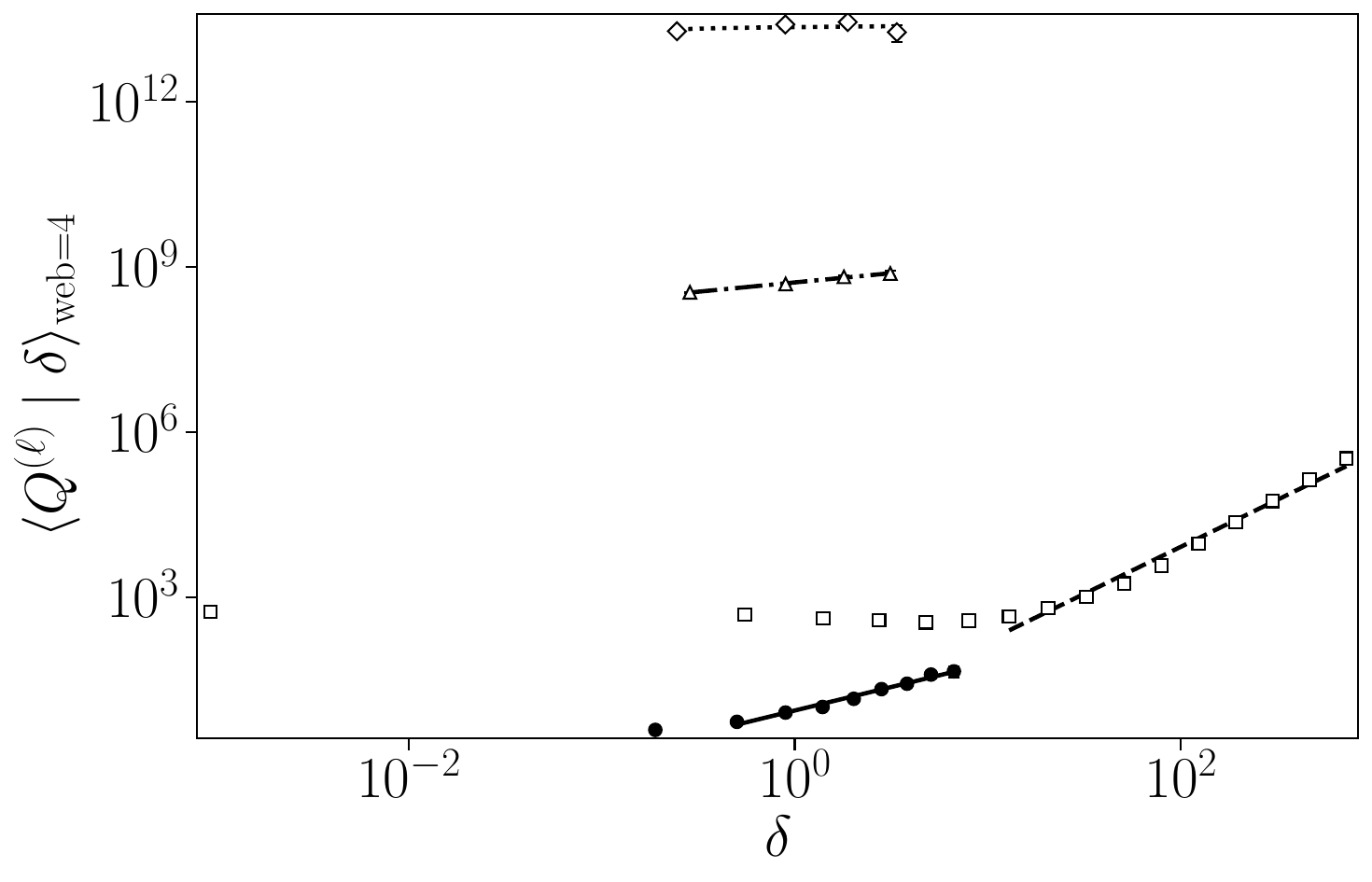}}
\put(-90,80){\footnotesize voids}
\end{tabular}
\vspace{-0.cm}
\caption{Same as Figure \ref{fig:landau-ginzburg} but shown separately for knots, filaments, sheets, and voids. The figure does not support a single universal hierarchy among web types. Instead, the anisotropic activation is strongly environment dependent. Sheets and filaments often display the clearest and most extended activated branches, while knots are more affected by strong nonlinearity and possible saturation effects. Voids are not simply featureless: for the lowest hierarchical levels they show some of the cleanest and most nearly linear trends, consistent with the idea that in regions with weaker multistreaming the minimal coarse-grained description can remain accurate over a broader density range.}
\label{fig:landau-ginzburg2}
\end{figure}

The shifted power-law fits presented in this appendix serve as empirical summaries of the full conditional anisotropy curves over broad density ranges. They are distinct from the restricted thresholded-linear fits used in the main text to connect the onset of anisotropy with the minimal Landau branch.
The conditional anisotropy curves shown in the main text are measured by binning the field uniformly in \(\log(1+\dd)\). If \(\mathcal B_n\) denotes the set of cells in the \(n\)-th bin, the measured conditional mean is
\begin{equation}
\bar Q_n^{(i)}
=
\frac{1}{N_n}\sum_{\vx\in\mathcal B_n}Q^{(i)}(\vx),
\label{eq:binned_Q_app}
\end{equation}
with standard error
\begin{equation}
\sigma_n^{(i)}
=
\frac{1}{\sqrt{N_n}}
\left[
\frac{1}{N_n-1}
\sum_{\vx\in\mathcal B_n}
\left(Q^{(i)}(\vx)-\bar Q_n^{(i)}\right)^2
\right]^{1/2}.
\label{eq:binned_Qerr_app}
\end{equation}

To characterize the activated branch, we fit a shifted power-law model,
\begin{equation}
\left\langle Q^{(i)} \mid \dd \right\rangle
=
C_i\,(\dd-\dd_{c,i})^{p_i},
\qquad
\dd>\dd_{c,i}.
\label{eq:shifted_powerlaw_app}
\end{equation}
For a fixed trial value of \(\dd_{c,i}\), taking logarithms gives a linear relation,
\begin{equation}
y_n^{\log}=\log \bar Q_n^{(i)}
=
\log C_i + p_i \log(\dd_n-\dd_{c,i}),
\label{eq:log_fit_app}
\end{equation}
where \(\dd_n\) is the representative density of the bin.

The best-fitting threshold \(\dd_{c,i}\) is selected by scanning trial values and minimizing the residual sum of squares in log-space,
\begin{equation}
{\rm SSE}_{\log}
=
\sum_n
\left(
y_n^{\log}-\hat y_n^{\log}
\right)^2.
\label{eq:SSE_log_app}
\end{equation}
A useful goodness-of-fit measure is the coefficient of determination in log-space,
\begin{equation}
R^2_{\log}
=
1-
\frac{\sum_n \left(y_n^{\log}-\hat y_n^{\log}\right)^2}
{\sum_n \left(y_n^{\log}-\bar y^{\log}\right)^2},
\label{eq:R2_log_app}
\end{equation}
with
\begin{equation}
\bar y^{\log}
=
\frac{1}{N_{\rm fit}}\sum_n y_n^{\log}.
\end{equation}
When binned standard errors are available, one may also compute a chi-square in log-space through the propagated uncertainties
\begin{equation}
\sigma_{n,\log}^{(i)}\simeq \frac{\sigma_n^{(i)}}{\bar Q_n^{(i)}},
\end{equation}
leading to
\begin{equation}
\chi^2_{\log}
=
\sum_n
\frac{
\left(y_n^{\log}-\hat y_n^{\log}\right)^2
}{
\left(\sigma_{n,\log}^{(i)}\right)^2
}.
\label{eq:chi2_log_app}
\end{equation}

The fitted parameters \(C_i\), \(p_i\), and \(\dd_{c,i}\) therefore provide, respectively, the amplitude, the growth slope, and the effective activation threshold of the anisotropic branch.
For completeness, we also examine the conditional anisotropy curves separately in different web environments. This decomposition is not central to the main free-energy argument, but it is useful for showing that the density--anisotropy relation is not universal across morphology classes. Instead, the activation pattern depends on environment. Sheets and filaments often exhibit the clearest and most extended activated branches, while knots are more strongly affected by nonlinear saturation and classification ambiguities. Table~\ref{tab:Q_fits} summarizes the best-fitting shifted power-law parameters for the conditional anisotropy relations, both for the full sample (Figure \ref{fig:landau-ginzburg}) and separately for knots, filaments, sheets, and voids (Figure \ref{fig:landau-ginzburg2}). Voids show a more nuanced behavior: at the lowest hierarchical levels they display some of the cleanest and most nearly linear trends, suggesting that in regions where multistreaming remains comparatively weak the minimal coarse-grained description can remain valid over a broader density range.

\begin{table*}
\caption{Best-fitting shifted power-law parameters for
\(\langle Q^{(\ell)}\mid\delta\rangle_\alpha\).}
\hspace{-1.cm}
\begin{tabular}{lcccccccccccc}
\toprule
& \multicolumn{3}{c}{$\ell=1$}
& \multicolumn{3}{c}{$\ell=2$}
& \multicolumn{3}{c}{$\ell=3$}
& \multicolumn{3}{c}{$\ell=4$} \\
\cmidrule(lr){2-4}\cmidrule(lr){5-7}\cmidrule(lr){8-10}\cmidrule(lr){11-13}
& $p$ & $\delta_c$ & $R^2_{\log}$
& $p$ & $\delta_c$ & $R^2_{\log}$
& $p$ & $\delta_c$ & $R^2_{\log}$
& $p$ & $\delta_c$ & $R^2_{\log}$ \\
\midrule
All
& 1.53 & 0.00 & 0.987
& 1.98 & 0.00 & 0.994
& 0.76 & 0.00 & 0.909
& 0.55 & 0.00 & 0.847 \\
Knots
& 1.36 & 0.00 & 0.962
& 1.77 & 55.95 & 0.999
& -1.04 & 0.00 & 0.966
& -0.61 & 0.00 & 0.573 \\
Filaments
& 1.43 & 0.00 & 0.963
& 1.91 & 0.00 & 0.993
& 1.10 & 0.00 & 0.957
& 0.85 & 0.00 & 0.964 \\
Sheets
& 1.56 & 0.00 & 0.998
& 1.42 & 0.00 & 0.964
& 1.31 & 0.00 & 0.982
& 1.03 & 0.00 & 0.938 \\
Voids
& 0.88 & 0.29 & 0.993
& 0.98 & 0.00 & 0.830
& 0.34 & 0.00 & 0.994
& 0.03 & 0.23 & 0.111 \\
\bottomrule
\end{tabular}
\label{tab:Q_fits}
\end{table*}

{\color{black}

\section{Initial spectra and the emergence of web-like structure}
\label{app:initial_spectra_web}

The emergence of complex web-like morphology requires more than the mere
presence of fluctuations. The initial field must contain fluctuations over a sufficiently broad range
of scales and must possess sufficient two-point spatial correlation so that
gradients and tidal deformations are coherent over extended regions. Exact
isotropy would lead only to locally spherical collapse, whereas generic
random fields possess unequal Hessian eigenvalues and therefore collapse
anisotropically. In this sense, web formation requires initial conditions
that seed both density contrast and eigenvalue splitting. Subsequent
transport, nonlocal coupling and gravitational or interaction-driven
amplification then convert these weak initial anisotropies into sheets,
filaments and knots.

A central role is played by the shape of the initial power spectrum. For a
statistically homogeneous field \(X\) in three spatial dimensions, the
dimensionless variance per logarithmic interval is
\begin{equation}
\Delta_X^2(k)
\equiv
\frac{k^3}{2\pi^2}P_X(k).
\end{equation}
Thus, if
\begin{equation}
P_X(k)\propto k^{-m},
\end{equation}
then
\begin{equation}
\Delta_X^2(k)\propto k^{3-m}.
\end{equation}
The exponent \(m\) therefore controls which scales dominate the initial
variance. The scale-invariant case in three dimensions corresponds to
\begin{equation}
m=3,
\end{equation}
for which each logarithmic interval in \(k\) contributes the same
dimensionless power.

In cosmology, this is the Harrison--Zel'dovich principle applied to the
primordial curvature perturbation \({\cal R}\):
\begin{equation}
\Delta_{\cal R}^2(k)
=
\frac{k^3}{2\pi^2}P_{\cal R}(k)
=
\mathrm{const},
\end{equation}
or equivalently
\begin{equation}
P_{\cal R}(k)\propto k^{-3}.
\end{equation}
For the matter density contrast, the corresponding large-scale
matter-era result is \(P_\delta(k)\propto k\), because
\(\delta\propto k^2\Phi\) while \(P_\Phi(k)\propto k^{-3}\).

The effect of deviations from this scaling can be understood directly from
\(\Delta_X^2(k)\propto k^{3-m}\). If \(m<3\), the spectrum is relatively
blue: the dimensionless power grows towards large \(k\), and the initial
variance is increasingly dominated by small scales. In such a case, the
first nonlinear objects tend to be smaller and more fragmented. The web
then contains more small-scale substructure, thinner filaments and more
numerous compact knots, provided that the small-scale power is not erased
by smoothing, pressure, diffusion, free streaming, or finite resolution.

If \(m>3\), the spectrum is relatively red: the dimensionless power grows
towards small \(k\), and the initial variance is dominated by large scales.
The resulting morphology is more coherent on large scales. Collapse is then
organized by long-wavelength modes, producing larger sheets, broader
filaments and more extended coherent patterns. If the spectrum is too red,
however, the field may become dominated by the largest modes available in
the finite volume, and the morphology can approach a few large coherent
features rather than a richly hierarchical web.

The Harrison--Zel'dovich case, \(m=3\), is intermediate. It gives no
preferred primordial scale in the dimensionless curvature perturbation:
large and small logarithmic intervals in \(k\) carry comparable initial
power. This does not by itself imply that all final structure sizes are
equally represented. The evolved matter field is shaped by the transfer
function, growth factor, free-streaming or pressure effects, baryonic
physics, nonlinear collapse and the finite smoothing scale of the
observation. Thus, even when the primordial curvature field is close to
scale invariant, the final matter distribution can acquire a characteristic
turnover and a scale-dependent hierarchy of structures.

The special case
\begin{equation}
P_X(k)=\mathrm{const}
\end{equation}
corresponds to \(m=0\). In three dimensions this gives
\begin{equation}
\Delta_X^2(k)\propto k^3.
\end{equation}
It is therefore strongly dominated by small-scale power. Such a spectrum is
white noise in the dimensional power spectrum, not scale invariance. By
itself, it does not naturally produce an extended coherent web. Instead it
tends to generate grainy, small-scale fluctuations with little long-range
phase coherence. A web-like pattern can still emerge after smoothing,
nonlocal transport or long-range interaction, but in that case the coherent
morphology is produced mainly by the subsequent dynamics or by the imposed
coarse-graining scale, rather than by scale-free initial correlations.

For complex web structures to emerge robustly, the initial conditions
should therefore satisfy several broad requirements. First, they should
contain fluctuations over a range of scales, rather than being concentrated
at a single wavelength. Second, they should include enough large-scale power
to generate coherent tidal or deformation fields. Third, they should also
retain sufficient small- and intermediate-scale power to allow
fragmentation, knots and internal filamentary substructure. Fourth, the
field should not be exactly isotropic locally: generic eigenvalue splitting
of the Hessian or deformation tensor is the seed of anisotropic collapse.
Finally, the spectrum should be processed by a dynamics capable of coupling
scales. A Gaussian field with a prescribed power spectrum supplies the
maximum-entropy initial baseline, but nonlinear transport, nonlocal
interaction and shell crossing are what convert that baseline into a
non-Gaussian web.

The same reasoning can be applied beyond cosmology, with the important
qualification that the measured field need not be a primordial curvature
perturbation. In another complex system, a measured \(P(k)\propto k^{-3}\)
over some finite range would indicate approximately scale-independent
dimensionless power in three dimensions over that range. It would not imply
that the microscopic dynamics is cosmological. Rather, it would suggest
that the system contains fluctuations distributed across scales in a way
that is statistically favourable to multiscale structure. Conversely,
spectra shallower than \(k^{-3}\) emphasize finer structures, while spectra
steeper than \(k^{-3}\) emphasize larger coherent structures.

Analogous spectral analyses have been used outside cosmology. For example,
Vazza and Feletti compared aspects of the neuronal network and the cosmic
web using spectral and network statistics \citep{Vazza20}. More recently,
cosmological power-spectrum and higher-order-statistics methods have been
applied to structural MRI data of the human brain
\citep{ROSELL2025121500,2025NatSR..1542309K}. These analogies should be
understood statistically rather than dynamically: similar approximate power-law behaviour over finite scaling ranges can
provide a common language  for multiscale organization, even
when the underlying microscopic mechanisms are entirely different.

}

\section*{Acknowledgements}

FSK thanks Rodolfo Cuerno Rejado, Francesco Sinigaglia for {\color{black} general} discussions, {\color{black}  Pau Amaro Seoane for detailed comments to the manuscript and Juan Reguera Vidaechea for discussions on the cytoskeleton}.  FSK also thanks  the Cabildo de Tenerife under the IACTEC Technological Training Program, grant TF INNOVA,  within the framework of MEDI-FDCAN 2016–2026, in support of the \textit{Cosmic Brain} project   \url{www.cosmic-brain.org}  (PI: FSK). FSK also thanks the \textit{Cosmology with Large Scale Structure Probes} project \url{www.cosmic-signal.org}  (PI: FSK) funded by the IAC; the Spanish Ministry of Science and Innovation for financing the \textit{Big Data of the Cosmic Web} project PID2020-120612GB-I00  (PI:FSK); and the \textit{FIRE (Field-level bayesian Inference to Reconstruct the univErse)} project PID2024-160504NB-I00  (co-PI:FSK).

\bibliographystyle{JHEP}

\bibliography{bibliography}

\end{document}